\documentclass[aps,prx,amsmath,twocolumn,amssymb,superscriptaddress,longbibliography]{revtex4-2}

\usepackage[colorlinks=true,citecolor=blue,linkcolor=magenta]{hyperref}
\usepackage{amssymb,amsfonts,amsmath}
\usepackage{color}
\usepackage[apple mac]{inputenc}
\usepackage[english]{babel}
\usepackage{times}
\usepackage{latexsym}
\usepackage{verbatim}
\usepackage{graphicx}
\usepackage{epsf}
\usepackage{subfigure}
\usepackage{bm}
\usepackage{amssymb}
\usepackage{stackrel}

\graphicspath{{./},{figures/},}

\usepackage{epstopdf}

\definecolor{Nathanblue}{rgb}{0.,0.24,0.51}
\newcommand{\blue}{\color{Nathanblue}}
\definecolor{orange}{rgb}{0.96,0.24,0.00}

\usepackage{color}

\usepackage{mathtools}

\def\be{\begin{equation}}
\def\ee{\end{equation}}

\begin{document}

\title{{\blue Engineering and probing non-Abelian chiral spin liquids using periodically driven ultracold atoms}}

%\title{{\blue Designing probes for the Kitaev chiral spin liquid using engineered periodically driven ultracold atoms}}

%\title{{\blue Engineering probes for the Kitaev chiral spin liquid using periodically driven ultracold atoms}}

%\title{{\blue Engineering and probing the Kitaev chiral spin liquid using periodically driven ultracold atoms}}

\author{Bo-Ye Sun}
\email[]{boyesun@ytu.edu.cn}
\affiliation{CENOLI,
Universit\'e Libre de Bruxelles, CP 231, Campus Plaine, B-1050 Brussels, Belgium}
\affiliation{YanTai University, Yantai, Shandong, 264005, People's Republic of China}

\author{Nathan Goldman}
\email[]{ngoldman@ulb.ac.be}
\affiliation{CENOLI,
Universit\'e Libre de Bruxelles, CP 231, Campus Plaine, B-1050 Brussels, Belgium}

\author{Monika Aidelsburger}
\email[]{monika.aidelsburger@physik.uni-muenchen.de }
\affiliation{Faculty of Physics, Ludwig-Maximilians-Universit\"at M\"unchen, Schellingstr.~4, D-80799 Munich, Germany}
\affiliation{Munich Center for Quantum Science and Technology (MCQST), Schellingstr.~4, D-80799 Munich, Germany}

\author{Marin Bukov}
\email[]{mgbukov@phys.uni-sofia.bg}
\affiliation{Max Planck Institute for the Physics of Complex Systems, N\"othnitzer Str.~38, 01187 Dresden, Germany}
\affiliation{Faculty of Physics, St.~Kliment Ohridski University of Sofia, 5 James Bourchier Blvd, 1164 Sofia, Bulgaria}

\begin{abstract}
We propose a scheme to implement Kitaev's honeycomb model with cold atoms, based on a periodic (Floquet) drive, in view of realizing and probing non-Abelian chiral spin liquids using quantum simulators.
We derive the effective Hamiltonian to leading order in the inverse-frequency expansion, and show that the drive opens up a topological gap in the spectrum without mixing the effective Majorana and vortex degrees of freedom. 
We address the challenge of probing the physics of Majorana fermions, while having only access to the original composite spin degrees of freedom.    
Specifically, we propose to detect the properties of the chiral spin liquid phase using gap spectroscopy and edge quenches in the presence of the Floquet drive. The resulting chiral edge signal, which relates to the thermal Hall effect associated with neutral Majorana currents, is found to be robust for realistically-prepared states. 
By combining strong interactions with Floquet engineering, our work paves the way for future studies of non-Abelian excitations and quantized thermal transport using quantum simulators.
\end{abstract}

\date{\today}

\maketitle

\section{\label{sec:intro}Introduction}

Quantum simulation is emerging as one of the most promising pillars of quantum technology, as it enables the observation of phenomena predicted to occur in models that are difficult to encounter in nature. A central ingredient for emulating the behavior of quantum systems is the ability to engineer the underlying Hamiltonian and probe its physics. Over the last decade, a toolbox was developed based on periodic (Floquet) drives, with the aim of imprinting novel physical properties by dressing the states of static systems~\cite{goldman2014periodically,bukov2015universal,eckardt2017colloquium}. 

This \textit{Floquet engineering} approach already proved instrumental for enabling the realization of artificial gauge fields and topological band structures in quantum simulators~\cite{goldman2016topological,eckardt2017colloquium,cooper2019topological,weitenberg2021tailoring}; however, applying it in the strongly-interacting regime remains an outstanding challenge at the forefront of present-day research, due to unwanted energy absorption from the drive. Strong quantum correlations are intrinsic to fractional quantum Hall states~\cite{leonard2022realization,clark2020observation}, symmetry-protected topological phases~\cite{de2019observation,sompet2022realizing}, spin liquids~\cite{verresen2021prediction,semeghini2021probing,satzinger2021realizing,xu2022digital,zhou2022probing}, and lattice gauge theories~\cite{martinez2016real,schweizer2019floquet,yang2020observation,mil2020scalable,banuls2020simulating,zhou2022thermalization}, some of which have been predicted to exhibit non-Abelian excitations~\cite{nayak2008non}. 
Understanding and harnessing the properties of topologically-ordered phases has far-reaching applications:~quantum computing codes, such as the toric code, make use of a degenerate ground state manifold for error-correction~\cite{kitaev2003fault,stern2013topological}; braiding of non-Abelian anyonic excitations (Majorana fermions and vortices) forms the basics of topological quantum computing~\cite{andersen2022observation,lensky2022graph,harle2022observing}; Majorana fermions exhibit chiral neutral currents associated with a quantized thermal Hall conductivity~\cite{read2000paired,kitaev2006anyons,nasu2017thermal,banerjee2018observation,kasahara2018majorana,grissonnanche2019giant,yamashita2020sample,yang2020universal,kapustin2020thermal,fang2021thermoelectric,bruin2022robustness}.  

\begin{figure}[t!]
\includegraphics[width = \linewidth]{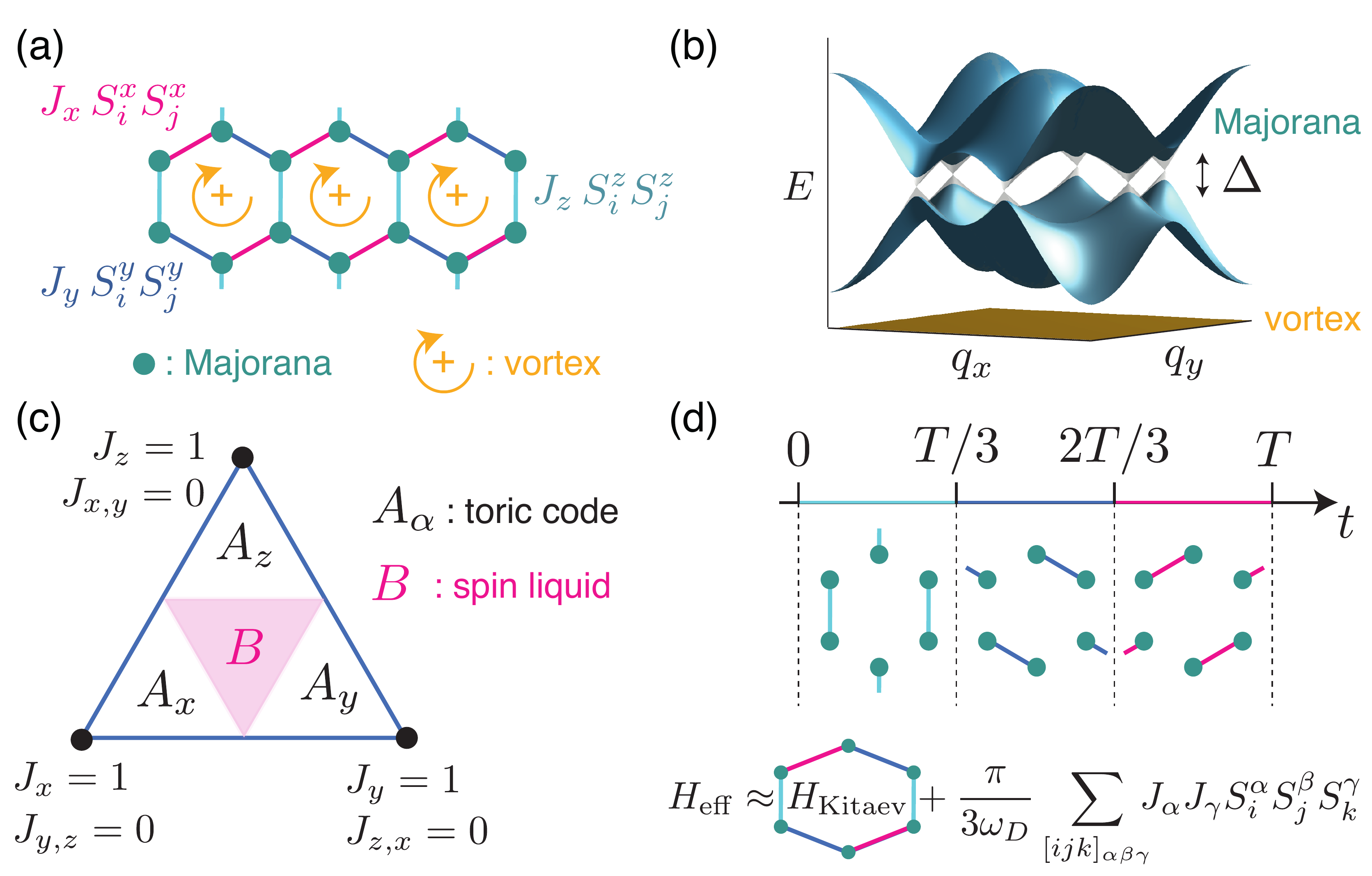}
\caption{
\textbf{Model and drive.}
\textbf{(a)} The Kitaev honeycomb model is an emblematic instance of a $\mathbb{Z}_2$ lattice gauge theory; it can be mapped exactly to Majorana fermions defined on the vertices of the honeycomb lattice (cyan), and dispersionless vortices residing on the plaquettes (orange). 
\textbf{(b)} Dispersion relation of the Majorana fermions (cyan bands) with gap closing at the Dirac cones. In our driving protocol, the vortex dispersion (orange) remains flat, which renders vortices immobile. 
\textbf{(c)} Ground state phase diagram of the original Kitaev model, featuring three disconnected gapped phases $A_\alpha$ (Abelian topological phases equivalent to the toric code), and a gapless spin-liquid phase $B$. 
\textbf{(d)} Floquet realization: periodically modulating the $x$, $y$, and $z$-bonds at frequency $\omega_D$ breaks time-reversal symmetry and opens up a topological gap $\Delta$ in the Majorana dispersion in phase $B$, leaving the vortex dispersion intact; see panel (b). The ground state of the effective Hamiltonian $H_\mathrm{eff}$ is a chiral spin liquid, a topological phase exhibiting non-Abelian excitations. 
}
\label{fig_1}
\end{figure}

A paradigmatic model that exhibits both Abelian and non-Abelian topological phases is the Kitaev honeycomb model~\cite{kitaev2006anyons}: 
\begin{equation}
    \label{eq:Kitaev_H}
    H_\mathrm{Kitaev} = -J_x\sum_{\langle i,j\rangle_x}  S^x_i S^x_j
        -J_y\sum_{\langle i,j\rangle_y}  S^y_i S^y_j
        -J_z\sum_{\langle i,j\rangle_z}  S^z_i S^z_j,
\end{equation}
where the spin-$1/2$ operators obey $[S^\alpha_i,S^\beta_j]{=}i\delta_{ij}\varepsilon^{\alpha\beta\gamma}S^\gamma_j$, $\alpha,\beta,\gamma{\in}\{x,y,z\}$, and we set $\hbar{=}1$. Each of the three distinct types of bonds on the honeycomb lattice (connecting nearest-neighboring sites $\langle i,j\rangle_{\alpha}$ along the $\alpha$ direction) carries an interaction of strength $J_\alpha$ along a single spin direction $\alpha$ [Fig.~\ref{fig_1}(a)]. 
The model is integrable and can be solved exactly by decomposing the spins into dispersive Majorana and dispersionless vortex degrees of freedom, which decouple completely~\cite{kitaev2006anyons} [Fig.~\ref{fig_1}(b)]. In the thermodynamic limit, the ground state features a uniform vortex configuration. 
The ground state phase diagram exhibits three disconnected gapped phases, labelled $A_\alpha$, where the low-energy physics of the system can be mapped to the toric code~\cite{kitaev2006anyons}. In between them, a gapless spin-liquid phase appears, marked $B$ in Fig.~\ref{fig_1}(c). Applying an external magnetic field breaks time-reversal symmetry and opens up a topological Majorana gap, giving rise to a chiral spin liquid; moreover, vortices acquire a dispersion and integrability is broken~\cite{kitaev2006anyons,gohlke2018dynamical}.

The Kitaev honeycomb model has spawned a flurry of interest in the solid-state community, triggering the search and study of the so-called Kitaev materials~\cite{trebst2022kitaev,hermanns2018physics,knolle2019field, takagi2019concept}. 
A smoking gun signature of chiral spin liquids is a half-quantized thermal Hall conductivity, similar to the quantized Hall effect in Chern insulators, but with neutral Majorana fermions as carriers (rather than electrons)~\cite{nasu2017thermal,joy2022dynamics}.
Different techniques have been proposed to dynamically probe the signatures of spin liquids~\cite{knolle2014dynamics,schmitt2015dynamical}, including neutron scattering~\cite{banerjee2017neutron}, quench protocols used to measure the spin structure factor~\cite{knolle2015dynamics,smith2016majorana,chen2018nonabelian,feldmeier2020local}, and to monitor edge dynamics of spinon wave-packets~\cite{mizoguchi2020oriented}.    

In this work, we propose a protocol to engineer the time-reversal-broken Kitaev honeycomb model using periodically-driven ultracold atoms. 
We further address the significant challenge of probing the topological properties associated with neutral, particle-nonconserving Majorana modes in quantum simulators, with only limited access to the spin degrees of freedom. To this end, we propose nonequilibrium protocols combining slow ramps with abrupt quenches in the presence of the Floquet evolution, to reveal the chiral spin liquid phase, and measure the edge heat currents as its hallmark signature. 
We close by discussing state preparation, opportunities for a  physical implementation, potential challenges and outlooks.

%%%%%%%%%%%%%%%%%

\section{\label{sec:F-engine}Floquet Engineering the Kitaev honeycomb model}

Consider a periodic drive that implements the following Floquet unitary
%~\cite{goldman2014periodically,bukov2015universal,eckardt2017colloquium} 
(i.e., the time-evolution operator over one period $T$ of the driving sequence): 
\begin{equation}
    \label{eq:KItaev_UF}
    U_F = 
    \mathrm e^{-i \frac{T}{3} J_x'\!\!\sum\limits_{\langle i,j\rangle_x}\!\!\!\!  S^x_i S^x_j  }\;
    \mathrm e^{-i \frac{T}{3} J_y'\!\!\sum\limits_{\langle i,j\rangle_y}\!\!\!\!  S^y_i S^y_j  }\;
    \mathrm e^{-i \frac{T}{3} J_z'\!\!\sum\limits_{\langle i,j\rangle_z}\!\!\!\!  S^z_i S^z_j  },
\end{equation}
where $T{=}2\pi/\omega_D$ defines the relation between the drive period and frequency; see Fig.~\ref{fig_1}(d). Physical implementations of this driving sequence, using quantum simulators, are discussed in Sec.~\ref{sec:implementation}.

A straightforward application of the van Vleck inverse-frequency expansion~\cite{goldman2014periodically,bukov2015universal,eckardt2017colloquium} shows that, in the high-frequency limit, the dynamics is stroboscopically generated by the effective Hamiltonian $H_\mathrm{eff}$ and kick operator $K_\mathrm{eff}$, according to
\begin{equation}
    U_F = \mathrm{e}^{-i K_\mathrm{eff}}\;\mathrm{e}^{-i T H_\mathrm{eff}}\; \mathrm{e}^{+i  K_\mathrm{eff}},
\end{equation}
with [see Appendix~\ref{app:IFE}]
\begin{eqnarray}
    \label{eq:H_eff}
    H_\mathrm{eff} &=& H_\mathrm{Kitaev}
    + \frac{\pi }{3\omega_D} \sum\limits_{ [ijk]_{\alpha\beta\gamma} } \!\!\! J_\alpha J_\gamma S^\alpha_i S^\beta_j S^\gamma_k {+}\mathcal{O}(\omega_D^{-2}),\\
    K_\mathrm{eff} &=&  -\frac{2\pi }{3\omega_D} \left( \sum\limits_{\langle ij\rangle_x} J_x S^x_i S^x_j {-} \sum\limits_{\langle ij\rangle_z} J_z S^z_i S^z_j \right) {+}\mathcal{O}(\omega_D^{-2})\; . \nonumber
\end{eqnarray}
Here $J_\alpha{=}{-}J_\alpha'/3$, and $[ijk]_{\alpha\beta\gamma}=\langle i,j\rangle_{\alpha}
\langle j,k\rangle_{\gamma}$ denotes three neighboring sites where all associated spin operators are different: $\alpha{\neq}\beta{\neq}\gamma$; see Fig.~\ref{fig_3body}. 
We point out that the kick operator is crucial for capturing the correct stroboscopic dynamics~\cite{goldman2014periodically,bukov2015universal,goldman2015periodically}.
While the infinite-frequency contribution to $H_\mathrm{eff}$ is readily recognizable as the Kitaev model [first term in Eq.~\eqref{eq:H_eff}], the $(1/\omega_D)$-correction opens up a chiral topological gap $\Delta$ between the two Majorana bands (at the Dirac point); see Fig.~\ref{fig_1}(b) and Fig.~\ref{fig_2}(a). Here, the gap magnitude is $\Delta{=}3\sqrt{3}g/4$, with $g{=}\pi J^2/(3\omega_D)$ for isotropic systems ($J_\alpha{=}J)$. Note that time-reversal symmetry is broken by the periodic drive; however, unlike applying an external magnetic field~\cite{kitaev2006anyons}, one can verify that vortices remain dispersionless under this Floquet drive to any order in the drive frequency [Fig.~\ref{fig_1}(b)], since the plaquette operators~\cite{kitaev2006anyons} commute with the generators of the three unitaries in Eq.~\eqref{eq:KItaev_UF}. Hence, $U_F$ preserves the vortex structure of the initial state, see Appendix~\ref{app:mapping}.

Related Floquet-Kitaev physics has recently been explored in Josephson junction arrays~\cite{sameti2019Floquet}, and in solid state systems~\cite{strobel2022comparing,kumar2022floquet}, setting the focus on anomalous topological phases of matter~\cite{fulga2019topology,fidkowski2019interacting,po2017radical}, and the characterization of edge modes and their properties~\cite{molignini2021crossdimensional}.

In the next Sec.~\ref{sec:probes}, we place the emphasis on the design of practical probes tailored for cold-atom simulators. Specifically, we first demonstrate the existence of an energy gap in the spectrum of $H_\mathrm{eff}$, using a practical spectroscopic approach. 
Then, we present a nonequilibrium protocol to probe the Majorana chiral edge transport, based on local spin-spin-correlation measurements. 
Finally, we discuss a state preparation sequence that allows us to demonstrate the topological properties of the system starting from a realistically-prepared state. 
We show numerical simulations of these protocols in the presence of the Floquet driving sequence that generates the model.

\section{\label{sec:probes}Probing Floquet-Kitaev physics}

Let us start by listing a number of outstanding issues, which are related to present-day limitations in measuring and probing chiral spin liquids using cold-atom simulators. (i) The Kitaev honeycomb model appears deceptively similar to the emblematic Haldane model of Chern insulators~\cite{haldane1988model,jotzu2014experimental,tarnowski2019measuring,asteria2019measuring}, when written in terms of Majorana fermions~\cite{kitaev2006anyons}; however, it is substantially more challenging to probe and analyze, since Majorana excitations are fractionalized excitations which do not conserve the number of particles. In particular, it is not possible to access the topological properties of this model through mass/particle transport measurements and traditional spectroscopic probes.
(ii) Majorana particles emerge as an effective degree of freedom:~in contrast, the fundamental degree of freedom, accessible on the quantum simulator, is the spin. 
Thus, the question arises as to which quantity to measure in order to observe the topological properties associated with the effective (Floquet-Kitaev) Hamiltonian. 
(iii) A different, yet equally important, issue concerns the state preparation: can one have access to topological signatures from a realistically-prepared state?
Last, (iv), we emphasize the critical role of the Floquet drive itself, which ensures that the vortex and Majorana degrees of freedom remain decoupled throughout the dynamics; this important feature of our scheme is instrumental for accessing the physics of chiral spin liquids in the quantum-simulation framework. 

To exclude extraneous features associated with a non-ideal initial state, we begin our analysis in 
Secs.~\ref{subsec:gap} and~\ref{subsec:edge_transport} by using the rigorously determined ground state of the system as the initial state for our proposed Floquet protocol. Subsequently, in Sec.~\ref{subsec:state_prep}, we examine the preparation of the initial state and its influence on the system's response.

\subsection{\label{subsec:gap}Gap spectroscopy}

\begin{figure}[t!]
\includegraphics[width = \linewidth]{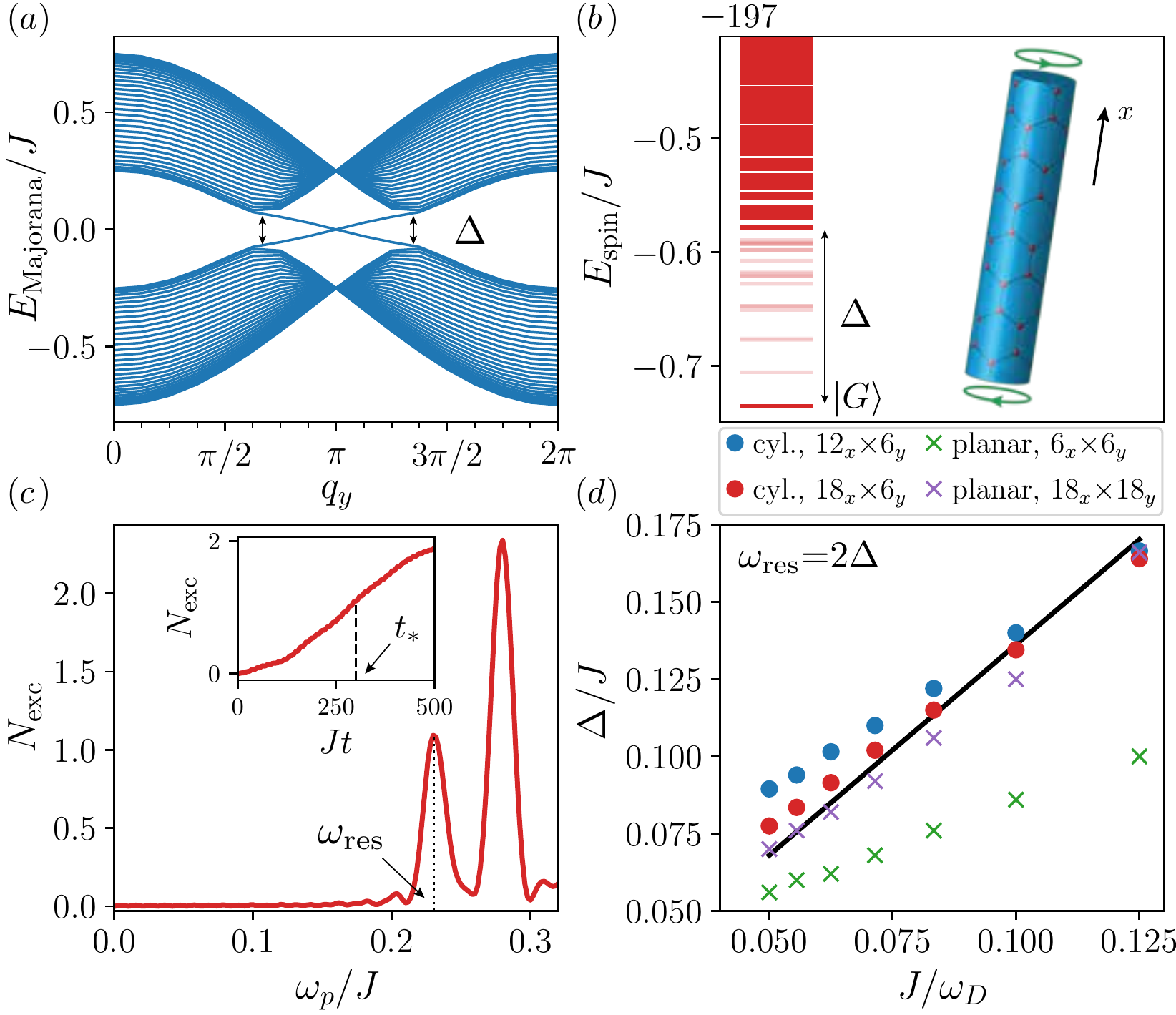}
\caption{
\textbf{Gap spectroscopy.}
\textbf{(a)} Majorana spectrum, $E_\mathrm{Majorana}$, showing a bulk gap opening at the Dirac points as well as chiral edge modes, for a system with periodic boundary conditions along the $y$-direction (cylindrical geometry).
\textbf{(b)} Many-body energy spectrum, $E_\mathrm{spin}$, of the spin Hamiltonian $H_\mathrm{eff}$ in the vortex-free sector, for a cylindrical geometry (inset). The many-body ground state $|G\rangle$ of the system is a product state of the filled lower Majorana band and the ground state of the vortex sector. The bulk many-body gap $\Delta$ is indicated, and the edge states are located within the corresponding shaded area.
\textbf{(c)} The number of excited particles, $N_\mathrm{exc}(t_\ast)$, shows well-defined resonances as a function of the probe frequency $\omega_p$ (dotted vertical line), from which the gap can be extracted: $\Delta{=}\omega_\mathrm{res}/2$. Inset: $N_\mathrm{exc}$ as a function of time for $\omega_p/J{=}0.23$; $t_\ast$ denotes the measurement time: $Jt_\ast{=}300$. Spectroscopy is performed in the presence of the Floquet drive on a $18_x{\times}6_y $ cylinder for $A_p{=}0.02$, $\omega_D/J{=}12$.
\textbf{(d)} The extracted gap decreases when increasing the drive frequency $\omega_D$, and shows good agreement with the theoretical value of the bulk gap, $\Delta{=}\pi\sqrt{3} J^2/(4\omega_D)$, of 
the static model $H_\mathrm{eff}$ [black line]. Circles (crosses) correspond to a cylindrical (planar) geometry. The legend shows the number of unit cells used in the calculations. 
}
\label{fig_2}
\end{figure}

We first discuss the detection of the bulk gap in the spectrum of the driven spin system [Eq.~\eqref{eq:KItaev_UF}], which opens as a consequence of the (chiral) periodic drive; see Fig.~\ref{fig_2}(b) and Section~\ref{sec:F-engine}. Since the Majorana fermions are emergent degrees of freedom, probing their dispersion directly poses a significant challenge for present-day quantum simulators. The difficulty arises from the presence of vortices to which an arbitrary probe would also generically couple.
Here, we address this question by showing how a practical spectroscopic probe, applied directly to the spin degrees of freedom, can accurately detect the gap opening.

As we show below, a spectroscopic probe can be simply realized through a sinusoidal modulation of the (uniform) coupling between the $z$-bonds, 
\begin{equation}
    \label{eq:gap_spectroscopy}
    J_z(t) {=} J_z\left(1 {+} A_p\sin\omega_p t \right),
\end{equation}
where $\omega_p$ and $A_p$ are the probe's frequency and amplitude, respectively.
In our simulations, we find that this protocol works properly in the time-scale separated regime $\omega_D^{-1}{\propto}\Delta{\sim}\omega_p{\ll}\omega_D$. 

Figure~\ref{fig_2}c (inset) shows the number of excited particles as a function of time, $N_\mathrm{exc}(t)$, starting from the ground state $|G\rangle$ and applying simultaneously the Floquet drive and the spectroscopic probe [Eq.~\eqref{eq:gap_spectroscopy}]. This quantity is proportional to the spectroscopic signal that one would detect in an experimental context. The main panel shows the dependence on the probe frequency $\omega_p$ at the measurement time $t_\ast$, chosen such that the resonant signal is detectable; in practice, this amounts to $Jt_\ast{\sim}10^2$. One observes clearly discernible resonant peaks corresponding to transition frequencies $\omega_\mathrm{res}$ between the ground and excited many-body states, with a peak positioned at $2\Delta$. The physical origin of this factor 2 is intimately related to the fractional character of the Majorana quasiparticles: it can be traced back to the restriction that excitations can only be created in pairs in the Kitaev model~\cite{kitaev2006anyons}. A rigorous proof for the appearance of this factor 2 can be found in Appendix~\ref{app:factor_2}.

In Fig.~\ref{fig_2}(d), we display the resonant frequency for different values of $\omega_D^{-1}$. Floquet theory [solid black line] predicts a gap closing in the infinite-frequency limit; see~Eq.~\eqref{eq:H_eff}. The extracted gap clearly follows this law for large enough systems, with deviations due to higher-order frequency corrections discernible at large values of $\omega_D^{-1}$. The numerical data shows a reasonable agreement for different system sizes, both for cylindrical (circles) and planar (crosses) geometries. In very small systems with boundaries (planar geometry), we note that finite-size effects can cause ambiguity in identifying the peak corresponding to the bulk gap, see Appendix~\ref{app:spetroscopy}. In Fig.~\ref{fig_2}(d), we explicitly compare different system sizes in order to help identify a suitable regime for a potential experimental realization.

\subsection{\label{subsec:edge_transport}Probing chiral Majorana edge transport}

\begin{figure}[t!]
\includegraphics[width = \linewidth]{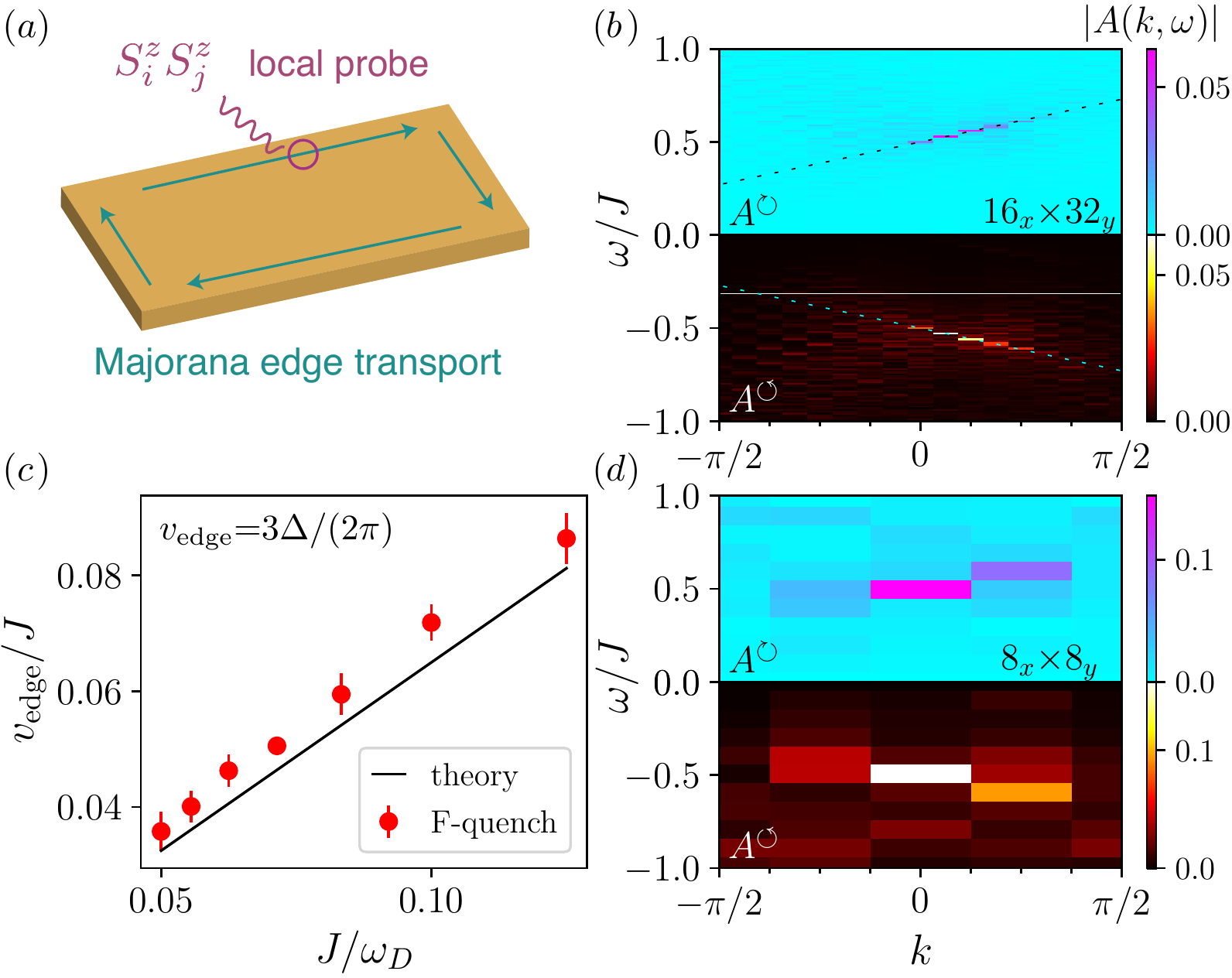}
\caption{
\textbf{Detecting chiral Majorana edge transport.}
\textbf{(a)} Schematic representation of the nonequilibrium quench protocol, which we use to reveal the chiral heat current in the ground state of $H_\mathrm{eff}$.
\textbf{(b)} The Fourier spectrum associated with $zz$-correlations on the edge, $C_m(\ell)$ [Eq.~\eqref{eq:quench_C}], which we use to probe the chiral dispersion relation of Majorana excitations. The slope in $(\omega,k)$ space [dashed line] equals half the group velocity of the edge mode excitations ($v_\mathrm{edge}$). Upper (cyan) and lower (red-hot) panels correspond to positive and negative circularly polarized Floquet drives: $A^\circlearrowright(k,\omega)$, $A^\circlearrowleft(k,\omega)$.
The system size is $16_x{\times}32_y$. 
The model parameters are $\omega_D/J{=}10$, $J_q/J{=}0.008$, and $N_T{=}1001$. 
\textbf{(c)} The group velocity of the chiral mode $v_\mathrm{edge}$ extracted from (b) decreases as one increases the Floquet drive frequency $\omega_D$, in agreement with Floquet theory (which predicts a vanishing edge-mode velocity in the infinite-frequency limit). The evolution times are proportional to the drive period, $JTN_T{\sim}100{\times}2\pi$ for every $\omega_D$-point. The error bars show the fit error, see Appendix~\ref{app:fitting_procedure}.
\textbf{(d)} Same as (b) but for a smaller system of size $8_x{\times}8_y$, and the remaining parameters are the same as in panel (b).
}
\label{fig_3}
\end{figure}

The detection of the bulk gap opening is insufficient to demonstrate the topological nature of the ground state associated with the effective time-reversal broken Kitaev Hamiltonian $H_\mathrm{eff}$. 
In chiral spin liquids, a natural signature of topological (non-Abelian) excitations is provided by the quantized thermal Hall effect~\cite{nasu2015thermodynamics,nasu2017thermal}.  This phenomenon, which arises from the chiral transport of heat along the edge of the system, was recently measured in ``Kitaev'' materials~\cite{kasahara2018majorana,yamashita2020sample,bruin2022robustness}.
Motivated by such energy transport measurements~\cite{nasu2015thermodynamics,nasu2017thermal}, we propose a non-equilibrium probe to detect the chiral transport of heat along the edge [Fig.~\ref{fig_3}a], and hence to reveal the chiral spin liquid nature of our Floquet-engineered system.

Let us consider a system with boundaries prepared in the ground state of the isotropic effective Hamiltonian $H_\mathrm{eff}$ (i.e.~$J_{\alpha}{=}J$).
We now perform a quench by adding a local perturbation on a single $z$-bond, $\langle i,j_z \rangle_z$, located on the edge of the system: $H_z{\to} H_z{+}J_q S^z_iS^z_{j_z}$; see Fig.~\ref{fig_3}a and Fig.~\ref{fig_def_edge}.
The modified Floquet unitary, which includes this local perturbation, is denoted by $\tilde U_{F}$. 
We let the system evolve under the quenched Floquet drive for a fixed number of cycles $\ell{\in}\{1,2,\dots,N_T\}$, and we measure the local excess of energy in every unit cell located on an edge:
\begin{eqnarray}
    \label{eq:quench}
    E_m(\ell) &{=}& \langle G|\left[\tilde U_{F}^\dagger\right]^\ell H_m \left[\tilde U_{F}\right]^\ell|G\rangle,    \nonumber\\
    H_m &{=}& \frac{J}{2} \sum_{i\in m} \left( S^x_iS^x_{j_x} + S^y_iS^y_{j_y} + S^z_iS^z_{j_z}\right) ,
\end{eqnarray}
where the unit cell $m$ is located on the edge, and $\langle i,j_x\rangle_{x}$, $\langle i,j_y\rangle_{y}$, $\langle i,j_z\rangle_{z}$ denote the sets of links connected to $m$; see~Fig.~\ref{fig_def_edge}.
Measuring $E_m(\ell)$ stroboscopically at each drive cycle, $\ell{\in}\{1,2,\dots,N_T\}$, gives rise to a timetrace of data.
The chiral nature of the edge dynamics can be revealed by post-processing this time-trace using a discrete Fourier transform~\cite{dong2018charge}, $A(k,\omega){=}\sum^{L_y}_{m=1}\sum_{\ell=1}^{N_T} E_m(\ell)\; \mathrm e^{-ik m +i \omega \ell T}$, where $L_y$ is the length of the edge considered. Indeed, we verified that this Fourier spectrum clearly captures the chiral dispersion relation of the Majorana edge modes; see Appendix~\ref{app:chiral_transport}.

A closer investigation reveals that the chiral signal is already detectable in the Fourier spectrum of the local $zz$-correlations on the edge:
\begin{equation}
    \label{eq:quench_C}
    C_m(\ell) {=} \langle G|\left[\tilde U_{F}^\dagger\right]^\ell J S^z_iS^z_{j_z} \left[\tilde U_{F}\right]^\ell|G\rangle,  
\end{equation}
where the edge $z$-bond $\langle i , j_z\rangle$ is shown in Fig.~\ref{fig_def_edge}.
This quantity is a constituent part of the local energy $E_m$ in Eq.~\eqref{eq:quench}; however, it requires fewer measurements. Therefore, for the sake of experimental simplicity, we focus our study on the behavior of $C_m$.
We note that this detection protocol requires unitary evolution (set by the modified Floquet sequence) and a single final measurement, which is particularly suitable for analog quantum simulators. In particular, it does not require the application of additional spin operators to create excitations on the edge of the sample, and whose effects have to be subsequently measured.

Figure~\ref{fig_3}(b),(d) shows the Fourier spectrum $|A(k,\omega)|$ extracted from $C_m(\ell)$, as computed from a numerical simulation of the  nonequilibrium (Floquet) quench, for two system sizes. In these simulations, we consider a cylindrical geometry, and we apply the quench protocol on opposite edges of the cylinder.
We observe a clear signal at low frequencies, which corresponds to the excitation of the chiral edge mode. 

Importantly, the slope of the chiral signal in $(\omega,k)$ space corresponds to twice the group velocity $v_\mathrm{edge}$ of the Majorana edge modes~\footnote{The edge mode velocity is measured in units of the distance $d$ between neighboring unit cells along the edge direction: $[v_\mathrm{edge}]{=}dJ/\hbar$.}; see Appendix~\ref{app:factor_2}.
We show the dependence of the extracted velocity on the Floquet drive frequency $\omega_D$ in Fig.~\ref{fig_3}(c); see Appendix~\ref{app:fitting_procedure} for details on the fitting procedure. In particular, we verify that the chiral signal disappears (i.e., the edge mode's velocity tends to zero) in the infinite-frequency limit, where the topological gap closes; see Eq.~\eqref{eq:H_eff}. 
Moreover, we have also verified that the chiral edge transport detected by our probe vanishes upon transitioning from the chiral spin liquid to the non-chiral phase of the phase diagram [Fig.~\ref{fig_1}(c)]. This confirms that the chiral edge signal is a genuine signature of the chiral spin liquid phase, and not a mere consequence of the circular ``polarization'' of the drive, cf~Fig.~\ref{figF2}.

Interestingly, one can use this measurement as an alternative to the spectroscopic measurement discussed in Sec.~\ref{subsec:gap}:~indeed, one can estimate the size of the topological bulk gap from the extracted velocity, using the relation $v_\mathrm{edge}{=}3\Delta/(2\pi)$.

Finally, to demonstrate a direct correspondence between the chirality of the edge mode and that of the Floquet drive, we reverse the ``polarization'', $\circlearrowright$, of the periodic driving sequence, and observe a reversal in the propagation direction of the excitations in the Fourier spectrum; compare the top and bottom panels in Fig.~\ref{fig_3}(b) and (d). A close examination shows that $|A^\circlearrowright(k,\omega)|{\neq}|A^\circlearrowleft(k,-\omega)|$; this implies that the Floquet quench excites a small fraction of non-chiral modes. Since these undesired excitations grow when increasing the quench strength $J_q$, it is preferable to work in the weak-quench regime $J_q/J{\ll}1$.

While the proposed measurement protocol is intimately related to the thermal Hall effect associated with non-Abelian excitations in chiral spin liquids~\cite{nasu2015thermodynamics,nasu2017thermal}, detecting the quantization of the thermal Hall conductivity remains an outstanding challenge in the context of quantum simulation; see Refs~\cite{brantut2013thermoelectric,husmann2018breakdown,salerno2019quantized,wu2022heat} regarding possible heat-transport measurements in quantum gases.

\subsection{\label{subsec:state_prep}State preparation sequence}

So far, we have shown how to detect hallmark features of the chiral spin liquid by assuming that the system can be initialized in the vortex-free ground state. However, preparing topological many-body states (which are entangled over long distances) from product states is a highly nontrivial task per se~\cite{bravyi2006lieb,konig2014generating,bespalova2021quantum,homeier2021Z2,tantivasadakarn2022hierarchy}. In particular, adiabatic quantum state preparation would require going through a topological phase transition, which is associated with a gap closing point in the thermodynamic limit. In quantum simulators, one can nevertheless exploit the finite size of the system (and hence, the possible existence of a finite-size gap at the topological phase transition) to perform adiabatic quantum state preparation~\cite{motruk2017phase,he2017realizing,leonard2022realization}.

To address this relevant aspect in detail, we build on existing ideas for preparing the ground state of the toric code~\cite{sahay2011quantum,petiziol2022nonperturbative,jin2022fractional}. 
At present, this is feasible using digital quantum simulators and requires the appropriate sequential application of high-fidelity Hardamard and CNOT gates~\cite{satzinger2021realizing}, or entangling gates combined with mid-circuit measurements~\cite{iqbal2023topological}; to the best of our knowledge, the optimal state preparation of the toric code ground state is currently an open problem in analog quantum simulators
\footnote{In addition, note that our small residual occupation of non-zero vortex sectors in the preparation of the toric code ground state is not expected to affect our state preparation protocol, since all stages of the dynamics preserve the vortex configurations.}.

\begin{figure}[t!]
\includegraphics[width = \linewidth]{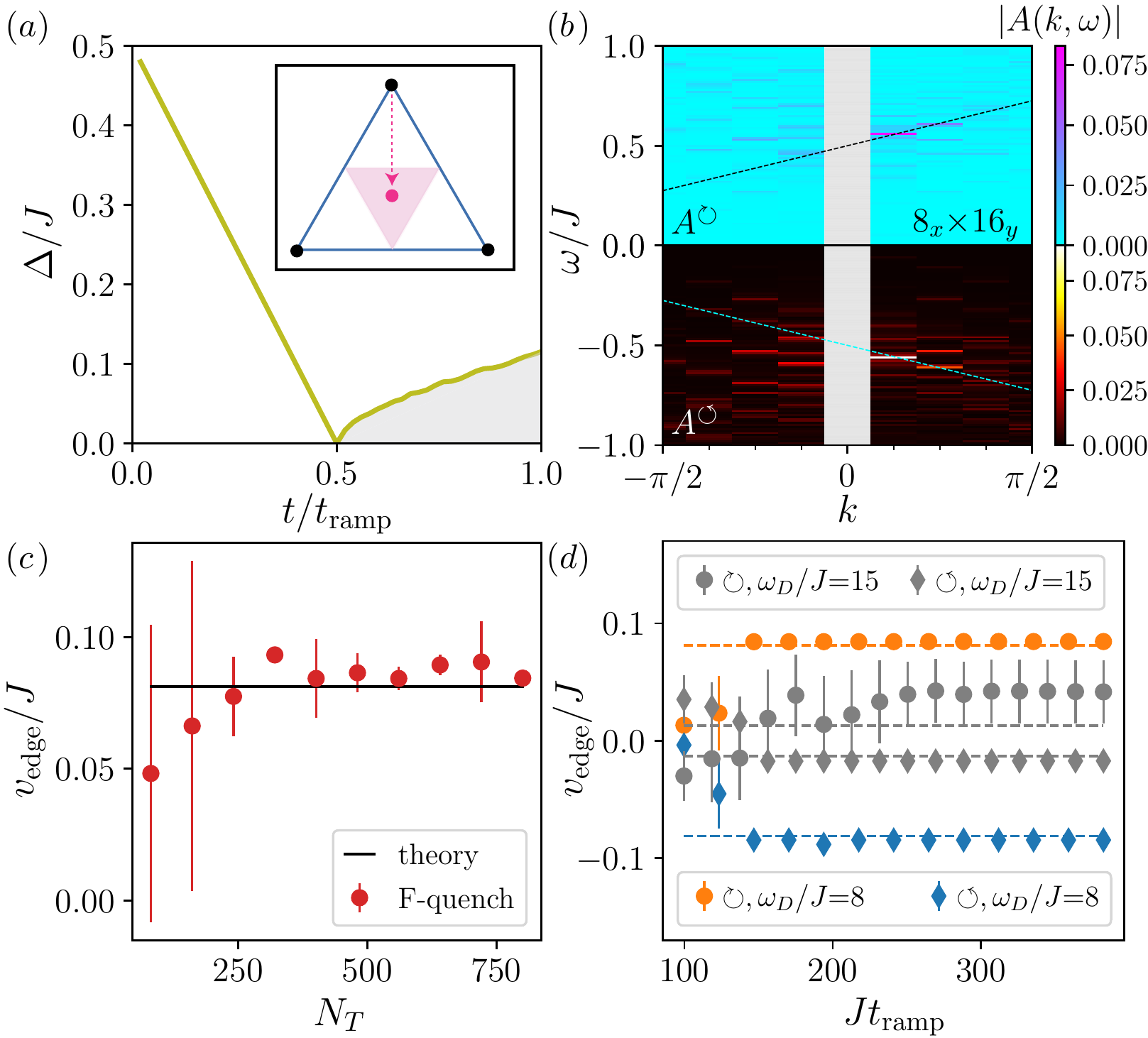}
\caption{
\textbf{Detecting chiral Majorana edge transport after quasi-adiabatic quantum state preparation.}
\textbf{(a)} The bulk Majorana gap at the Dirac point closes at the critical point between phases $A_z$ and $B$, during the ramp $J_\alpha(t){=}Jt/t_\mathrm{ramp}$, $\alpha{=}x,y$ (inset, see also Fig.~\ref{fig_1}c). The grey shaded area marks the existence of edge modes in the chiral spin liquid phase.
\textbf{(b)} Same as Fig~\ref{fig_3}(b), but with the initial state for the quench $U_F{\to}\tilde U_F$ prepared from the ground state at $J_x{=}J_y{=}0$ using the ramp protocol in (a). The grey region at $k{=}0$ shows a strong signal, disregarded in the postprocessing procedure. Chiral velocity fits (dashed straight lines) are performed in the regions $(k{>}0,\omega{>}0)_\circlearrowright$ and $(k{>}0,\omega{<}0)_\circlearrowleft$. The parameters are $t_\mathrm{ramp}{=}250J$, $N_T{=}1001$, $\omega_D{=}10J$. 
\textbf{(c)} The chiral current velocity $v_\mathrm{edge}$, extracted from the fits in (b), as a function of the observation cycle number $N_T$ after the quench (details can be found in Appendix~\ref{app:fitting_procedure}). Longer observation times allow to better resolve the signal and reduce the fit error. The parameters are $Jt_\mathrm{ramp}{=}250$, and $\omega_D{=}8J$.
\textbf{(d)} Chiral mode velocity fits $v_\mathrm{edge}$ as a function of the ramp duration $t_\mathrm{ramp}$ for two drive frequencies $\omega_D/J{=}8,15$. Horizontal dashed lines correspond to the theory values. 
%The required ramp durations to detect the chiral signal are within experimentally accessible times.
The number of observation cycles is $N_T{=}801,1501$. The system size is $8_x{\times}16_y$ for all panels.
}
\label{fig_4}
\end{figure}

The toric code ground state describes the low-energy physics of our system in the close vicinity of the point $J_{x,y}{=}0$~\cite{kitaev2006anyons}; see Fig.~\ref{fig_1}(c). 
Thus, starting from this (Abelian topological) gapped state, we can gradually cross the critical point~\cite{motruk2017phase} and end up in the desired (non-Abelian) gapped state, in the presence of the Floquet drive. This amounts to slowly ramping-up the couplings: 
\begin{equation}
    \label{eq:ramp}
    J_{x,y}(t) {=} J t/t_\mathrm{ramp},\qquad J_z=J,
\end{equation}
over a time $t_\mathrm{ramp}{\gg}T$, while keeping $J_z{=}J$ fixed. We note that the ramp terminates at the isotropic point $J_\alpha{=}J$, and that the bulk gap closes at the critical point [Fig.~\ref{fig_4}a], which inevitably causes excitations. Nevertheless, we find that the latter are not detrimental to the chiral edge signal introduced in Sec.~\ref{subsec:edge_transport} for realistic finite system sizes, as we now discuss.

We apply the ramp~\eqref{eq:ramp} in the presence of the evolution generated by $U_F(J_\alpha(t))$ for a duration $t_\mathrm{ramp}$, and then perform the quench $U_F{\to}\tilde U_F$ following the protocol described in Sec.~\ref{subsec:edge_transport}. 
Figure~\ref{fig_4}(b) shows the corresponding Fourier spectrum $A(k,\omega)$, which displays a noticeable chiral signal. Besides, one clearly observes additional (undesired) bulk excitations, which are created due to the finite ramp duration.

We point out that, even in the ideal ``adiabatic" limit where $t_\mathrm{ramp}{\to}\infty$, a long observation time $t_\mathrm{obs}{=}N_TT$ is required for accurately extracting the chiral signal from the Fourier spectrum. This is illustrated in Fig.~\ref{fig_4}(c), which shows the convergence of the extracted edge mode velocity as one increases the observation time.

Moreover, we emphasize that the finite duration of cold-atom experiments limits the duration of the ramp  $t_\mathrm{ramp}$ used to prepare the state of interest. Figure~\ref{fig_4}(d) shows that a ramp duration of the order of $Jt_\mathrm{ramp}{\approx}150$ suffices to clearly detect the chiral signal ($v_\mathrm{edge}{\ne}0$), as we illustrate by comparing the results obtained for Floquet sequences of opposite chirality [circle vs.~diamond markers]. We verify that the edge mode velocity decreases when increasing the drive frequency, $v_\mathrm{edge}{\propto}\omega_D^{-1}$, in agreement with theory; see the grey datapoints in Fig.~\ref{fig_4}(d) and Fig.~\ref{figA7}.

We stress that the ramp time required for adiabatic quantum state preparation necessarily diverges in the thermodynamic limit. Nevertheless, our simulations demonstrate that clear signatures of the chiral spin liquid should be detectable in realistic system sizes and within reasonable time scales.

\section{\label{sec:implementation}Physical implementation}

Having discussed the detection of topological signatures associated with the effective Hamiltonian $H_\mathrm{eff}$, let us now turn to the details of its physical implementation. We argue that it is possible to either use ultracold fermions (with Hubbard interactions), dipolar interactions between molecules or Rydberg-atom-based spin systems for this purpose.

A first cold-atom realization of the Kitaev model (not relying on high-frequency periodic drives) was proposed in Ref.~\cite{duan2003controlling}. However, this proposal remains challenging to implement in practice, since it requires a large number of laser beams and a two-photon Raman coupling. Moreover, it remains unclear whether it can be generalized to open up the chiral topological gap, while preserving the vortex configuration.

\subsection{\label{subsc:Floquet_impl}Implementation of the Floquet drive}

Consider the spinful Fermi-Hubbard model on a honeycomb lattice:
\begin{eqnarray}
	\label{eq:H_Hubbard}
	H_\mathrm{Hubbard} &{=}& \sum_{\substack{\langle ij\rangle, \\ \sigma}}\!\!\left({-}J_\sigma c^\dagger_{i\sigma}c_{j\sigma}  {+} \mathrm{h.c.} \right)  
	  {+}  U \sum_{j}  n_{j\uparrow}n_{j\downarrow},
\end{eqnarray}
where $J_\sigma$ denotes the strength of nearest-neighbor species-dependent hopping, and $U$ is the onsite interaction strength. The fermionic operators obey the anti-commutation relation $\{c_{i\sigma},c^\dagger_{j\sigma'}\}{=}\delta_{ij}\delta_{\sigma\sigma'}$.
Recall that, in the strongly-interacting atomic limit, $J_\sigma\ll U$, double occupancies are suppressed, and the spectral function exhibits gaps of order $U$, separating the so-called Hubbard bands.
In this regime, the system is described by the Heisenberg model~\cite{duan2003controlling,greif2013short,jepsen2020spin,bohrdt2021exploration}:
\begin{equation}
	\label{eq:H_Heisenberg}
	H_\mathrm{Heisenberg} {=} {-}\sum_{\langle ij\rangle} J_{zz} S^z_i S^z_j {+}\frac{J_\perp}{2}\left( S^+_i S^-_j {+} \mathrm{h.c.}\right), 
\end{equation}
where $J_{zz} {=} {-} 2(J_\uparrow^2 {+} J_\downarrow^2)/U $ is the Ising interaction, and
$J_\perp {=} {-} 4 J_\uparrow J_\downarrow/U$ is the exchange interaction. The spin operators are defined in terms of the fermion operators as
\begin{eqnarray}
	S^z_j {=} \frac{1}{2}\left( n_{j\uparrow} - n_{j\downarrow} \right), \quad
	S^+_j {=} c^\dagger_{j\uparrow}c_{j\downarrow},\quad
	S^-_j {=} c^\dagger_{j\downarrow}c_{j\uparrow}.
\end{eqnarray}

Freezing one of the spin species (i.e., suppressing maximally its hopping matrix element, e.g., $J_\downarrow{=}0$) inhibits the exchange interactions, $J_\perp{=}0$, and leaves only the global nearest-neighbor Ising $z$-interaction $J_{zz}$. Alternatively, an Ising-only interaction can be achieved by applying a magnetic field gradient to suppress the exchange interactions, and then turning on a weak resonant periodic drive to enable the $z$-interaction term~\cite{yuao_2011_controlling}; we remark that the latter is a resonant nonequilibrium scheme, which may produce unwanted heating, when combined with the primary Floquet drive from Eq.~\eqref{eq:KItaev_UF}.  
Recently, it was also shown that a controllable anisotropy in the Heisenberg couplings can be realized using a system of Rydberg arrays~\cite{scholl2022microwave}, or by means of Floquet engineering in superconducting qubits~\cite{nguyen2022programmable} and ultracold molecules~\cite{christakis2022probing}.

\begin{figure*}[t!]
\includegraphics[width = \linewidth]{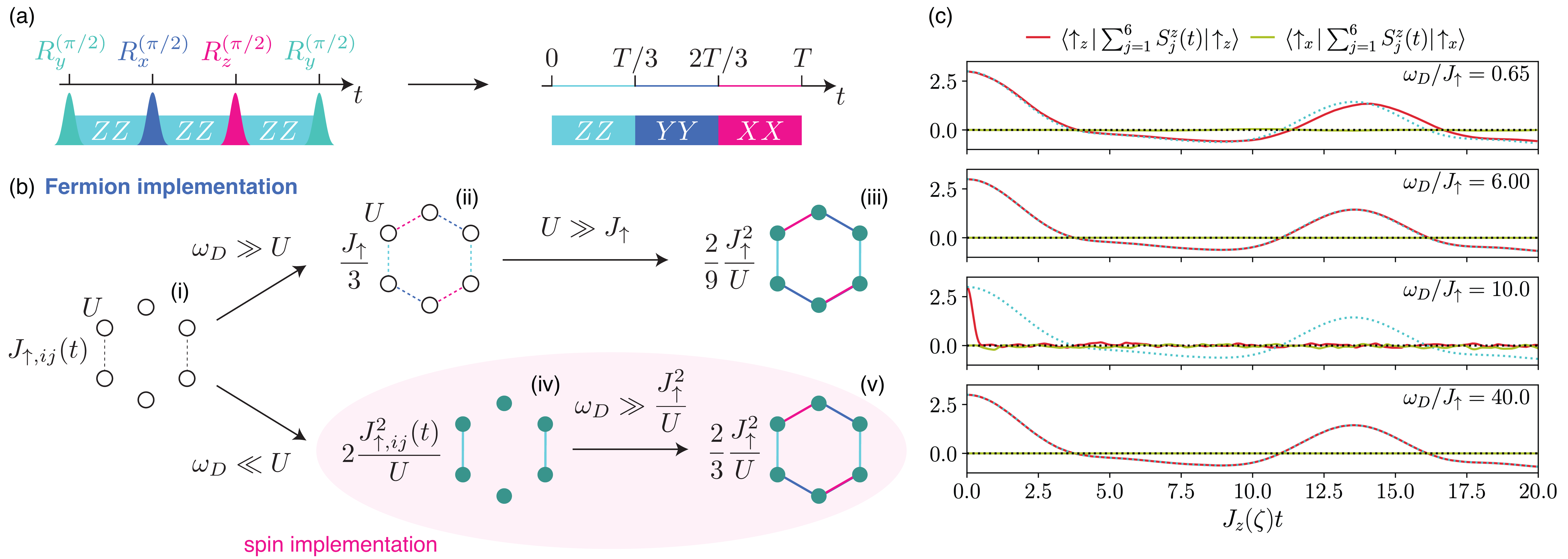}
\caption{
\textbf{Physical implementation.}
\textbf{(a)} Short $\pi/2$-pulses in between the steps of the Floquet drive, applied on a timescale much shorter than the inverse coupling strength $1/J$, turn the $z$-interactions into $x$ and $y$-interactions.
\textbf{(b)} Fermion implementation of the Floquet drive: (i) Fermi-Hubbard model with periodic modulation of the hopping, cf.~Eq.~\eqref{eq:Jup(t)}. In the high-frequency limit, $J_\uparrow{\ll}U{\ll}\omega_D$, applying the inverse-frequency expansion (i)$\to$(ii) leads to a modified static fermion Hamiltonian with nearest-neighbor hopping terms (also in the spin channel) of strength $J_\uparrow/3$. A Schrieffer-Wolff transformation (iii) then gives the effective spin-$1/2$ Hamiltonian $H_\mathrm{eff}$ from Eq.~\eqref{eq:H_eff} with interaction strength $J{=}2J_\uparrow^2/(9U)$, which governs the physics of the lower Hubbard band stroboscopically.
Alternatively, (i)$\to$(iv), in the strongly-interacting limit $J_\uparrow{\ll}\omega_D{\ll}U$, we first apply the Schrieffer-Wolff transformation to arrive at the spin-$1/2$ Hamiltonian~\eqref{eq:spin_impl}, with periodic in time, spatially oscillating $z$-interaction $J_{zz,ij}(t){=}2J^2_{\uparrow,ij}(t)/U$. A subsequent application of the inverse-frequency expansion then gives rise to $H_\mathrm{eff}$  with interaction strength $J{=}2J_\uparrow^2/(3U)$.
Step (iv) can also be independently taken as the starting point for the implementation in a spin-1/2 quantum simulator (red shaded oval).
\textbf{(c)} Time evolution of the $z$- and $x$-magnetization on a single honeycomb cell (open boundary conditions), for a few different values of the drive frequency $\omega_D/J_\uparrow$. The initial states are product states along the spin $z$- and $x$-direction (see legend).
The fermionic Floquet implementation (solid lines) agrees well with the dynamics generated by the spin-only Hamiltonian $H_\mathrm{eff}$ (dotted lines) over a wide range of drive frequencies $\omega_D$, except in the vicinity of resonances where $l\omega_D{\approx}U$ ($l{\in}\mathbb{N}$).
The model parameters are $J_\downarrow/J_\uparrow{=}0$ and $U/J_\uparrow{=}20$, with $J_z(\zeta)$ given in Eq.~\eqref{eq:Jeff}.
}
\label{fig_5}
\end{figure*}

With this idea in mind, the basic steps behind the fermionic implementation of the Kitaev model can be summarized as follows: 
First, we eliminate the spin-exchange term, which ensures that each bond interacts along a single spin direction (Ising-$z$).
%see~Fig~\ref{fig_5}b (i)$\to$(iv).
Second, we apply a periodic three-step modulation of the hopping matrix element $J_\uparrow(t)$ to isolate the different spatial bond types ($x$, $y$, and $z$) in time. %[Fig~\ref{fig_5}b (iv)$\to$(v)].
Finally, strong single-particle $\pi/2$-pulses in the spin channel, applied between the steps of the periodic drive, rotate the $z$-interactions into the desired type ($x$ or $y$). %[Fig~\ref{fig_5}(a)].
Applying the periodic step drive at a moderately high drive frequency results in the Hamiltonian from Eq.~\eqref{eq:H_eff}. 

Let us now explain the procedure in more detail. In the first step, a time-periodic spatial modulation of the non-zero hopping matrix element, $J_\uparrow(t{+}T){=}J_\uparrow(t)$, can be used to single out one type of bonds per Floquet step [see~Fig.~\ref{fig_1}(d)]:
\begin{equation}
    \label{eq:Jup(t)}
	J_{\uparrow,ij}(t) {=}\! \left\{  
	\begin{array}{lll}
		\!\! J_{\uparrow,ij}, &  \langle ij\rangle {\in} \{ \text{$z$-bonds} \}, & t\!\!\!\! \mod T {\in}[0,\frac{T}{3} 	) \\
		\!\! J_{\uparrow,ij}, &  \langle ij\rangle {\in} \{ \text{$y$-bonds} \}, & t\!\!\!\! \mod T  {\in}[\frac{T}{3},\frac{2T}{3} )   \\ 
		\!\! J_{\uparrow,ij}, &  \langle ij\rangle {\in} \{ \text{$x$-bonds} \}, & t\!\!\!\! \mod T  {\in}[\frac{2T}{3},T)\; .  
	\end{array}
\right.
\end{equation}
This can be achieved using a drive similar to Ref.~\cite{wintersperger2020realization}. In this scheme the strength of the tunnel coupling along the three different directions is modulated as a function of time by changing the intensities of the individual laser beams that form the underlying honeycomb lattice. The main challenge for experimental realizations will be to implement such a scheme with state-dependent optical potentials, that enable a state-dependent manipulation of the tunnel couplings. In principle, such a drive can also be implemented with quantum gas microscopes~\cite{gross2021quantum} using tightly-focused optical tweezer beams, whose position and strength can be controlled dynamically, as recently demonstrated~\cite{spar_realization_2022,yan_two-dimensional_2022}. Again, the main challenge lies in the realization of the required state dependence.
The interaction strength $U$ is kept constant on all bonds throughout the drive. 
This periodic modulation gives rise to a piece-wise constant periodic Hubbard Hamiltonian $H_\mathrm{Hubbard}(t) {=} (H_\mathrm{Hubbard}^z, H_\mathrm{Hubbard}^y, H_\mathrm{Hubbard}^x)$; the superscript $\alpha{=}x,y,z$ here refers to the spatial bonds in the honeycomb lattice [see~Fig.~\ref{fig_5}b (i)]. In the atomic limit, $J_\uparrow{\ll}U$, all terms $H_\mathrm{Hubbard}^\alpha$ give rise to a nearest-neighbor Ising $z$-interaction on the $x,y,z$-links, respectively.

At this stage, in the limit $J_\uparrow{\ll}U$, the system is effectively described by a spin-$1/2$ Ising model, whose Ising interaction is periodically modulated to switch between the different bonds on the honeycomb lattice [see Fig.~\ref{fig_5}b (iv)]: 
\begin{equation}
    \label{eq:spin_impl}
    H(t){=}{-}\sum_{\langle ij\rangle}J_{zz,ij}(t)S^z_iS^z_j,\qquad J_{zz,ij}(t){=}{-}\frac{2J^2_{\uparrow,ij}(t)}{U}.
\end{equation}
Incidentally, Eq.~\eqref{eq:spin_impl} can also be taken as the starting point for a spin-$1/2$ implementation  [Fig.~\ref{fig_5}(b), red oval], e.g., using Rydberg simulators~\cite{kalinowski2022non} operated in the nearest-neighbor interaction regime~\cite{hollerith2022realizing}. 

To generate the Floquet unitary in Eq.~\eqref{eq:KItaev_UF}, it remains to apply global on-site $\pi/2$-rotations in the spin channel; if they occur at a time scale much faster than the inverse interaction strength $J_{zz}$, this will consecutively rotate the $z$- into $y$- and $x$-interactions; see~Fig.~\ref{fig_5}(a). Due to the Euler angles theorem, it suffices to implement $\pi/2$ pulses along two out of the three spin axes. 

Notice that the discussion so far tacitly assumed the limit $J_\uparrow{\ll}\omega_D{\ll}U$, which places the drive frequency in between the two lowest Hubbard bands. Hence, the above analysis is equivalent to first applying a Shrieffer-Wolff transformation (SWT) [to eliminate the largest energy scale $U$, see Fig.~\ref{fig_5}b (i)$\to$(iv)], followed by the inverse-frequency expansion (IFE) [second-largest scale $\omega_D$]; see Fig.~\ref{fig_5}b (iv)$\to$(v). The effective Ising interaction is then given by $J_{z}{=}J_{zz}/3{=}{-}2J_\uparrow^2/(3U)$. 
By contrast, in the high-frequency limit, $J_\uparrow{\ll}U{\ll}\omega_D$, we have to first apply the IFE [Fig.~\ref{fig_5}b (i)$\to$(ii)] and then the SWT, which leads to $J_{z}{=}J_{zz}/9{=}{-}2J_\uparrow^2/(9U)$ [Fig.~\ref{fig_5}b (ii)$\to$(iii)]. Note that the Ising interaction strength $J_{z}$ is larger~\cite{mentink2015ultrafast} -- by a factor of three -- in the regime $J_\uparrow{\ll}\omega_D{\ll}U$, due to a dressing by the Floquet drive, as compared to the high-frequency regime. 
We note that these effective couplings are related to those from Eq.~\eqref{eq:KItaev_UF} via $J_z{=}{-}J'_z/3$, etc.

Setting $\zeta{=}\omega_D/U$, the two limits can be nicely unified by the more general expression, valid away from resonances $U{=}l\omega_D$ ($l{\in}\mathbb{N}$):
\begin{equation}
	\label{eq:Jeff}
	J_{z}(\zeta) = -\frac{2J^2_\uparrow}{U} 
	\sum\limits_{\substack{\ell=-\infty\\ \zeta\ell \neq -1}}^\infty \left( \frac{ \sin(\ell\pi/3) }{\ell\pi}\right)^2\frac{1}{1+\ell\zeta},
\end{equation}
which is obtained by reconciling the SWT and the IFE into a single framework~\cite{bukov2016SWT}, see Appendix~\ref{app:IFE}. 

Figure~\ref{fig_5}(c) shows that the stroboscopic magnetization dynamics of the Fermi-Hubbard model on a single honeycomb cell, following the above Floquet protocol [solid lines], agrees well with the dynamics of $H_\mathrm{eff}$ [dotted lines], for a wide range of \textit{non-resonant} frequencies~\footnote{The dynamics of $H_\mathrm{eff}$ is simulated in the smaller spin Hilbert space.}. Notice the proper, $\zeta$-renormalized scaling of the $x$-axis, which is essential for capturing the correct effective $z$-interaction across two orders of magnitude in the drive frequency (top to bottom panels). 

Finally, we note that the strongly-interacting regime is particularly appealing for Floquet engineering, since the condition $J_\uparrow{\ll}U$ readily ensures the high-frequency limit w.r.t.~the Majorana bandwidth of the Kitaev model:
$2J_\uparrow^2/(3U){\propto}J{\ll}\omega_D{\ll}U$. Thus, the proper high-frequency regime to observe Kitaev physics in the lower-Hubbard band is set by $2J_\uparrow^2/(3U){\ll}\omega_D$. In particular, it follows that even frequencies $\omega_D{\lesssim}J_\uparrow$ may yield reasonably good agreement with the dynamics of the target effective Hamiltonian [see Fig.~\ref{fig_5}(c), top-most panel], so long as no direct resonances are hit.

\subsection{\label{subsc:impl_probes}Implementation of the probing protocols}

The anisotropy in the spin couplings can be controlled by setting the durations of the corresponding periodic pulses, respectively. 
Thus, in platforms where modulating $J_z$ periodically in time is infeasible, the spectroscopic drive from Sec.~\ref{subsec:gap} can be effectively implemented by slowly changing in time the duration of only the $z$-bonds pulselength in the Floquet unitary~\eqref{eq:KItaev_UF}.
The same idea can be used to implement the slow ramp in Sec.~\ref{subsec:state_prep}; this will also enable the study the phase transition between phases $A$ and $B$ in the time-reversal broken Kitaev model. 
Finally, note that to realize the local quench on selected $z$-bonds from Sec.~\ref{subsec:edge_transport}, one may use a quantum gas microscope~\cite{bakr2009quantum,sherson2010single,gross2021quantum} to imprint the perturbation; alternatively, if the atoms are captured in tweezers with controllable positions~\cite{yan_two-dimensional_2022}, one could change the bond distance locally, which will affect the corresponding interaction strength.

\subsection{\label{eq:impl_challenges}Potential challenges}

The most prominent challenge for a potential experimental realization arises from the need to stabilize the global on-site spin rotations out to a large number of driving cycles. However, this could be potentially solved by adding spin-echo pulses to cancel dephasing, as recently demonstrated with molecules~\cite{christakis2022probing}. 
A second issue may occur due to the necessity to keep the temperature of the state at the order of, or below, the topological gap, which is only a fraction of the superexchange energy scale. Ultracold fermionic systems have recently reached low enough temperatures to observe antiferromagnetism~\cite{boll2016spin,mazurenko2017cold,brown2017spin} in the half-filled Fermi-Hubbard model. However, to reach low enough temperatures in order to observe ordered states, even lower temperatures are needed. The expected energy scales are of similar order as compared to the ones that appear in the current work~\cite{bohrdt_exploration_2021}. There are several promising techniques that have been proposed in order to reach these temperature scales~\cite{chiu_quantum_2018,kantian_dynamical_2018}. and we anticipate that despite this significant challenge cooling techniques will continue to improve. This will be beneficial for a number of fermionic quantum simulation experiments. Moreover, as we have shown above, detecting the chiral energy currents using nonequilibrium probes can still be robust to a certain amount of excitations. Finally, the gap size is controlled by the drive frequency; we observed numerically that $\omega_D{\approx} 8J$ is about the smallest drive frequency for which resonances within the Majorana bands are still suppressed.

\section{\label{sec:outro}Concluding remarks}

An important issue inherent to the Floquet engineering of strongly-interacting systems concerns detrimental heating processes~\cite{jin2022fractional}. While a comprehensive study of energy absorption and entropy creation in this nonequilibrium setting is beyond the scope of this work, we expect thermalization to infinite-temperature in the driven spin system to be absent for frequencies above the single-particle Majorana bandwidth: this follows from the conservation of the plaquette quantum numbers under the Floquet drive, which renders the Floquet Hamiltonian effectively single-particle, and makes the present model particularly appealing and promising for quantum simulation.
That said, secondary heating effects can still be caused by the presence of resonant processes with higher bands, or due to the creation of doubly occupied sites in the fermionic realization. 

Looking forward, two exciting directions concern (i) the investigation of local operations that would allow for the braiding of vortices and Majorana degrees of freedom (both in the bulk and on the edge)~\cite{kraus2012preparing,andersen2022observation,lensky2022graph,harle2022observing}; having dispersionless (i.e., immobile) vortices at our disposal is expected to prove instrumental in this context;
(ii) the design of probes in view of demonstrating the half-quantization of the thermal Hall conductivity associated with chiral edge transport.
Going beyond the present effective Kitaev Hamiltonian $H_\mathrm{eff}$, we note that it is well within the scope of our Floquet engineering protocol to incorporate long-range interactions, or even add an additional Heisenberg term to the effective Hamiltonian; this will enable the quantum simulation of a larger class of Hamiltonians describing Kitaev materials. 
Thus, by analyzing genuinely nonequilibrium protocols for probing and engineering topological Hamiltonians, our work paves the way for investigating Floquet topological phases of matter without static counterparts~\cite{po2017radical,fidkowski2019interacting} using quantum simulators, and demonstrates the possibility to reconcile Floquet engineering protocols with strong interactions to study strongly-correlated quantum phases of matter. 
\\   

%\newpage
%%%%%%%%%%%%%

\textit{Note added:} While finalizing this manuscript, we became aware of a similar proposal based on Rydberg digital quantum simulation~\cite{kalinowski2022non}.

\textit{Acknowledgments:} We acknowledge discussions with C.~Hickey, C.~Laumann, R.~Moessner, F.~Pollmann, C.~Repellin, and L.~Vanderstraeten. Work in Brussels is supported by the FRS-FNRS (Belgium), the EOS program (CHEQS project) and the ERC (TopoCold and LATIS projects).
M.B.~was supported by the Marie Sk\l{}odowska-Curie grant agreement No 890711. B.Y.S.~was supported by the China Scholarship Council and National Natural Science
Foundation of China (Grants No. 12204399). 
M.A.~acknowledges funding from the
Deutsche Forschungsgemeinschaft (DFG, German Research Foundation) via Research Unit FOR 2414 under project number 277974659, and under Germany's Excellence Strategy -- EXC-2111-390814868.

\textit{Author contributions:} M.B. and N.G. conceived the research, with inputs from B.Y.S. and M.A. B.Y.S and M.B. performed the numerical simulations. All authors analyzed the results and contributed to the writing of the manuscript. M.B. and N.G. supervised the work on the project.

\appendix

\section{\label{app:IFE}Details of the Floquet engineering}

To derive Eq.~\eqref{eq:H_eff} from Eq.~\eqref{eq:KItaev_UF}, we use the van Vleck inverse-frequency expansion (IFE)~\cite{goldman2014periodically,bukov2015universal,eckardt2017colloquium}. Floquet's theorem postulates that, for a time-periodic Hamiltonian $H(t)=H(t+T)$, $T=2\pi/\omega_D$, the time-evolution operator factorizes into a product of unitaries:
\begin{eqnarray}
    U(t,0)&=&\mathcal{T}\mathrm e^{-i\int_0^t\mathrm{d}sH(s)} \nonumber\\
    &=& \mathrm e^{-iK_\mathrm{eff}(t)}\; \mathrm e^{-itH_\mathrm{eff}}\; \mathrm e^{+iK_\mathrm{eff}(0)},
\end{eqnarray} 
where $H_\mathrm{eff}$ is the time-independent Floquet Hamiltonian and $K_\mathrm{eff}(t)=K_\mathrm{eff}(t+T)$ is the time-periodic generator of micromotion, also known as the kick operator. In the van Vleck IFE, $K_\mathrm{eff}$ satisfies the boundary condition $\int_0^T\mathrm{d}t K_\mathrm{eff}(t)=0$, which ensures that $H_\mathrm{eff}$ does not depend on the phase of the Floquet drive [by contrast, the Floquet-Magnus expansion uses the Floquet gauge $K_\mathrm{F}(t')=0$ for some $t'\in[0,T)$].

\begin{figure}[h!]
\includegraphics[width = 0.67\linewidth]{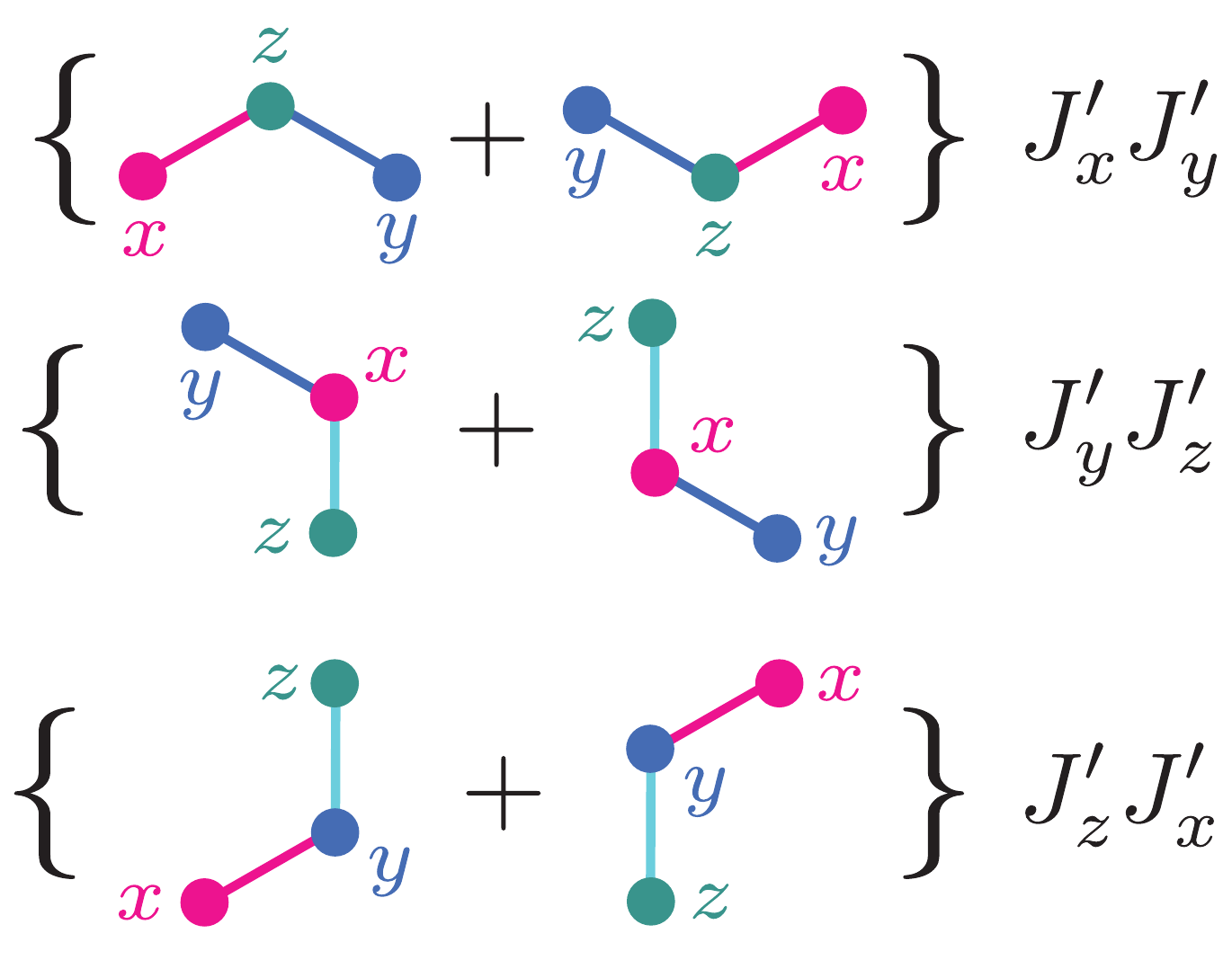}
\caption{
Schematic representation of the notation, $[ijk]_{\alpha\beta\gamma}$, showing the three-body terms to order $\omega^{-1}_D$, which are responsible for opening the topological bulk gap. The letters refer to the spin operator applied to each vertex, and the bond colors correspond to the bonds in the Kitaev honeycomb model; see ~Fig~\ref{fig_1}(a).
}
\label{fig_3body}
\end{figure}

Consider again the Floquet unitary from Eq.~\eqref{eq:KItaev_UF}:
\begin{eqnarray}
    U_F &=& 
    \mathrm e^{-i \frac{T}{3} J_x'\!\!\sum\limits_{\langle i,j\rangle_x}\!\!\!\!  S^x_i S^x_j  }\;
    \mathrm e^{-i \frac{T}{3} J_y'\!\!\sum\limits_{\langle i,j\rangle_y}\!\!\!\!  S^y_i S^y_j  }\;
    \mathrm e^{-i \frac{T}{3} J_z'\!\!\sum\limits_{\langle i,j\rangle_z}\!\!\!\!  S^z_i S^z_j  }. \nonumber
\end{eqnarray}
In this section, we will use the primed interaction strength $J'_\alpha$, which is related to the one used in the main text by $J_\alpha = -J'_\alpha/3$. The Floquet unitary is generated by the time-periodic Hamiltonian 
\begin{eqnarray}
    H(t) &=& -c\left(t\right)\sum_{\langle i,j\rangle_z}  J'_z S^z_i S^z_j \nonumber\\
        &&-c\left(t-\frac{T}{3}\right)\sum_{\langle i,j\rangle_y}  J'_y S^y_i S^y_j \nonumber\\
        && -c\left(t-\frac{2T}{3}\right)\sum_{\langle i,j\rangle_x}  J'_x S^x_i S^x_j,
\end{eqnarray}
with the piece-wise constant, time-periodic three-step function
\begin{equation}
    c(t) = \left\{
    \begin{array}{cc}
         &\!\!\! 1,\qquad 0\leq t\;\mathrm{mod}\; T \leq T/3  \\
         &0, \qquad T/3\leq t\;\mathrm{mod}\; T \leq T. 
    \end{array}
    \right.
\end{equation}

We can expand the effective Hamiltonian and kick operator to leading-order in the inverse-frequency $\omega_D$, 
\begin{eqnarray}
    H_\mathrm{eff}&=&H_\mathrm{eff}^{(0)}+H_\mathrm{eff}^{(1)} + \mathcal{O}\left(\omega_D^{(-2)}\right), \nonumber\\
    K_\mathrm{eff}&=&K_\mathrm{eff}^{(1)} + \mathcal{O}\left(\omega_D^{(-2)}\right),
\end{eqnarray}
using the weighted time-ordered integrals:
\begin{widetext}
\begin{eqnarray}
    H_\mathrm{eff}^{(0)}&=&  \frac{1}{T}\int_0^T\mathrm{d}t\,H(t) = H_0 \nonumber\\
    H_\mathrm{eff}^{(1)}&=& \frac{1}{2iT}\int_{0}^{T}\mathrm{d}t_1\int_{0}^{t_1}\mathrm{d}t_2\, \left[\left(1 - 2\frac{t_1-t_2}{T} \right)\; \mathrm{mod}\; 2\right] [H(t_1),H(t_2)] =\frac{1}{\omega_D}\sum_{\ell=1}^\infty \frac{1}{\ell} [H_\ell,H_{-\ell}],   \\
    K_\mathrm{eff}^{(1)}(t)&=&  -\frac{1}{2}\int_{t}^{T+t}\mathrm{d}t'\left[ \left(1 + 2\frac{t-t'}{T}\right)\; \mathrm{mod}\; 2\right]H(t')
    = \frac{1}{i\omega_D}\sum_{\ell\neq 0}\frac{\mathrm{e}^{i\ell\omega_D t}}{\ell} H_\ell,
\end{eqnarray}
\end{widetext}
where we fix the convention $(x\;\mathrm{mod}\; 2 ){\in}[-1,1)$. The right-hand side equations make use of the Fourier expansion of the Hamiltonian, given by $H(t)=\sum_{\ell\in\mathbb{Z}}\mathrm e^{+i\ell\omega t}H_\ell$.

\begin{figure}[t!]
\includegraphics[width = \linewidth]{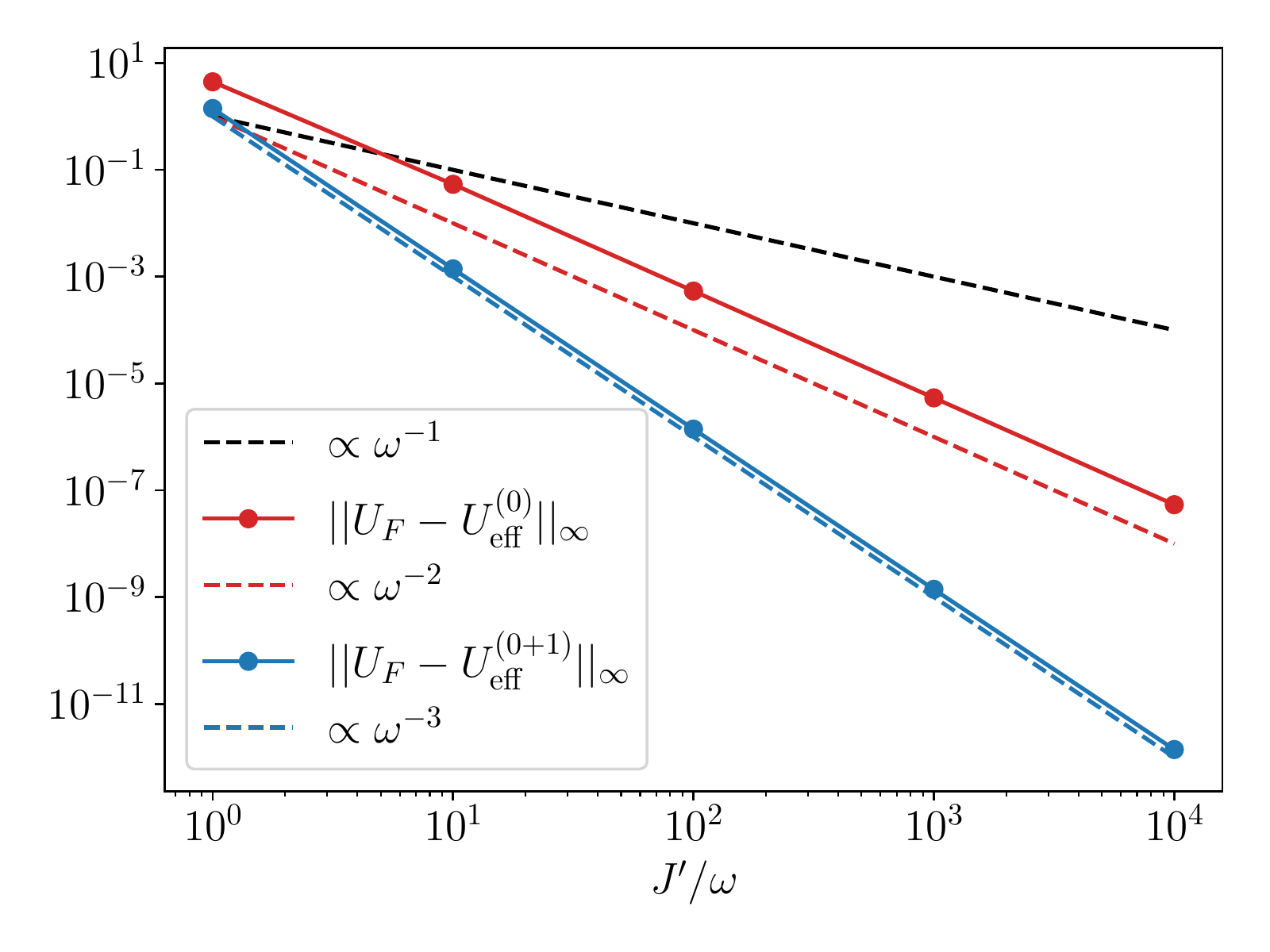}
\caption{
Numerical test of the effective Hamiltonian and kick operator against an exact simulation of a spin-$1/2$ system. Here, 
$U_\mathrm{eff}^{(0)} {=} \exp(-iTH_\mathrm{eff}^{(0)})$, and
$U_\mathrm{eff}^{(0{+}1)} {=} \exp(-iK_\mathrm{eff}^{(1)}(0))\exp(-iT[H_\mathrm{eff}^{(0)}{+}H_\mathrm{eff}^{(1)}])\exp(+iK_\mathrm{eff}^{(1)}(0))$.
The data show clearly that all corrections to the desired order are captured by Eq.~\eqref{eq:H_eff}. Note that the transformation generated by the kick operator is crucial for the validity of the test.
The system size is a single honeycomb cell, and all interactions are isotropic $J'_\alpha=J'$.
}
\label{fig_IFE}
\end{figure}

For the spin-$1/2$ implementation from Eq.~\eqref{eq:KItaev_UF}, we readily obtain (with $J=-J'/3$):
\begin{widetext}
\begin{eqnarray}
    H_\mathrm{eff}^{(0)} &=& 
        \frac{1}{3}\left(
        \sum_{\langle i,j\rangle_x}  J'_x S^x_i S^x_j
        +\sum_{\langle i,j\rangle_y}  J'_y S^y_i S^y_j
        +\sum_{\langle i,j\rangle_z}  J'_z S^z_i S^z_j
        \right), \nonumber \\
    H_\mathrm{eff}^{(1)} &=& \frac{\pi }{27\omega_D} \sum\limits_{ [ijk]_{\alpha\beta\gamma} } \!\!\! J'_\alpha J'_\gamma S^\alpha_i S^\beta_j S^\gamma_k,\qquad \text{[cf.~Fig.~\ref{fig_3body}]} \\
    % K_\mathrm{eff}^{(1)}(t) &=&  \frac{2\pi}{3\omega_D} \left( 
    % k(t) \sum\limits_{\langle ij\rangle_x} J_x S^x_i S^x_j 
    % + k\left(t+\frac{2T}{3}\right)\sum\limits_{\langle ij\rangle_y} J_y  S^y_i S^y_j 
    % + k\left(t+\frac{T}{3}\right) \sum\limits_{\langle ij\rangle_z} J_z  S^z_i S^z_j \right),\nonumber
    K_\mathrm{eff}^{(1)}(t) &=&  -\frac{2\pi}{9\omega_D} \left(
    k\left(t\right) \sum\limits_{\langle ij\rangle_z} J'_z  S^z_i S^z_j
    + k\left(t-\frac{T}{3}\right)\sum\limits_{\langle ij\rangle_y} J'_y  S^y_i S^y_j
    +k\left(t-\frac{2T}{3}\right) \sum\limits_{\langle ij\rangle_x}
    J'_x S^x_i S^x_j 
     \right),\nonumber
\end{eqnarray}
\end{widetext}
with the piece-wise linear $T$-periodic function 
\begin{equation}
    % k(t) = \left\{
    % \begin{array}{cc}
    %      & 6t/T-1,\qquad 0\leq t\;\mathrm{mod}\; T \leq T/3  \\
    %      \!-\!\!\!& 3t/T+2, \qquad T/3\leq t\;\mathrm{mod}\; T \leq T 
    % \end{array}
    % \right.
    k(t) = \left\{
    \begin{array}{cc}
         \!-\!\!\!& 6t/T+1,\qquad 0\leq t\;\mathrm{mod}\; T \leq T/3  \\
         & 3t/T-2, \qquad T/3\leq t\;\mathrm{mod}\; T \leq T. 
    \end{array}
    \right.
\end{equation}
The three-body terms in $H_\mathrm{eff}^{(1)}$ encompassed by the notation $[ijk]_{\alpha\beta\gamma}$ are shown in Fig.~\ref{fig_3body}. In turn, by investigating the scaling with $\omega_D$, in Fig.~\ref{fig_IFE} we show that the above expressions for the effective Hamiltonian and kick operator are complete.

In the fermion implementation, additional terms can arise due to higher-order terms of the Schrieffer-Wolff transformation, of order $\mathcal{O}(U^{-2}), \mathcal{O}(U^{-1}\omega_D^{-1})$,
$\mathcal{O}(\omega_D^{-2})$, etc. We leave their investigation to a future study. 

Instead, here we focus on the derivation of the effective interaction-renormalized Ising coupling, cf.~Eq.~\eqref{eq:Jeff}. To this end, we apply the generalized Schrieffer-Wolff transformation (SWT): using similar arguments to the derivation of Eq.~(31) in the Supplemental Material to Ref.~\cite{bukov2016SWT}, we pretend that the drive is resonant with the interaction, $U=l\omega_D$ ($l\in\mathbb{N})$, and do analytic continuation away from the resonance in the end. In particular, the effective Ising coupling in the generalized SWT is given by
\begin{eqnarray}
    J_z &=& -2\frac{J_\uparrow^2}{\omega_D}
    \sum\limits_{\substack{\ell=-\infty\\ \ell \neq 0}}^\infty
    \frac{|c_{\ell-l}|^2}{\ell}
    = -2\frac{J_\uparrow^2}{\omega_D}
    \sum\limits_{\substack{\ell=-\infty\\ \ell \neq -l}}^\infty
    \frac{|c_{\ell}|^2}{l+\ell}\nonumber\\
    &=& -2\frac{J_\uparrow^2}{U}
    \sum\limits_{\substack{\ell=-\infty\\ \ell \neq -l}}^\infty
    \frac{|c_{\ell}|^2}{1+\ell/l}
    =-2\frac{J_\uparrow^2}{U}
    \sum\limits_{\substack{\ell=-\infty\\ \ell\omega_D \neq -U}}^\infty
    \frac{|c_{\ell}|^2}{1+\ell \omega_D/U} \nonumber \\
    &=&-2\frac{J_\uparrow^2}{U}
    \sum_{\ell=-\infty}^\infty
    \frac{|c_{\ell}|^2}{1+\ell \omega_D/U}
    \label{eq:Jeff_derivation}
\end{eqnarray}
where $c_\ell$ are the Fourier coefficients of the Floquet drive $c(t)$, and we used the relation $l=U/\omega_D$ a few times. Notice that the last equality is only valid sufficiently far away from resonance, i.e., for $U\neq l\omega_D$.

Using the Fourier coefficients for the periodic three-step drive,
\begin{equation}
    c_\ell = \mathrm e^{-i\ell\frac{\pi}{3}}\frac{\sin\left(\ell{\pi}/{3}\right)}{\ell\pi},
\end{equation}
gives the final expression
\begin{equation}
    J_z\left(\frac{\omega_D}{U}\right) = -2\frac{J_\uparrow^2}{U}
    \sum_{\ell=-\infty}^\infty
    \left(\frac{\sin\left(\ell{\pi}/{3}\right)}{\ell\pi}\right)^2\frac{1}{1+\ell \omega_D/U}.
\end{equation}
In the infinite-frequency limit, $U\ll\omega_D$, only the $\ell=0$ harmonic survives, which leads to $J_z = -2J_\uparrow^2/(9U)$. By contrast, in the strongly-interacting limit, $\omega_D\ll U$, all harmonics contribute equally; using the relation $\sum_{\ell=-\infty}^{\infty} \left(\frac{\sin\left(\ell\pi/3\right)}{\ell\pi}\right)^2 {=}1/3$, then leads to $J_z = -2J_\uparrow^2/(3U)$. 

Since the derivation in Eq.~\eqref{eq:Jeff_derivation} is, to a certain degree, ambiguous, and given the heuristic character of the analytic continuation, we performed numerical tests, shown in Fig.~\ref{fig_5}c, which demonstrate the validity of Eq.~\eqref{eq:Jeff} sufficiently far away from resonance.

\section{\label{app:mapping}Mapping the Kitaev Hamiltonian to Majorana and vortex degrees of freedom}

In this Appendix, we provide more details on the representation of the Kitaev honeycomb model in terms of Majorana fermions, which plays a central role in this work. We focus on the isotropic case $J_\alpha=J$, but the generalization to the anisotropic model is straightforward.

\begin{figure}[t!]
\includegraphics[width = 0.8\linewidth]{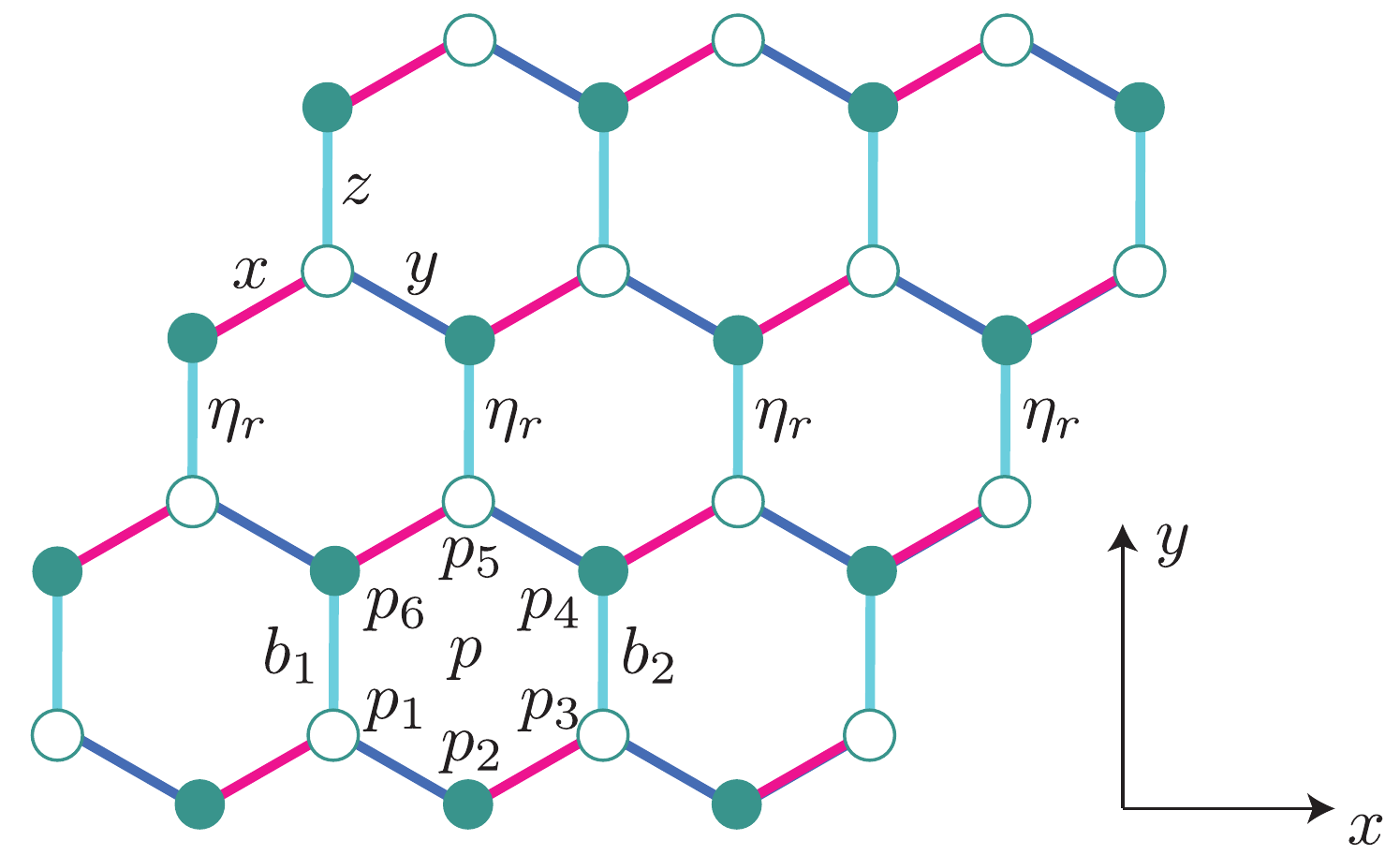}
\caption{Schematic representation of the notations used in Eq.~\eqref{MajH}. The $p$-th plaquette $p$ contains the six lattice sites $p_1,\dots,p_6$, with $b_1$ and $b_2$ denoting the two $z$-bonds on the plaquette. The unit cell contains sites from two sublattices marked by open and filled circles. $x$, $y$, and $z$ refer to the  direction of the bounds in the Hamiltonian.
}
\label{figA3}
\end{figure}

For the time-reversal broken Kitaev model, the effective Hamiltonian used in the main text is~\cite{kitaev2006anyons} 
\begin{equation}
    H=-J\sum_{\alpha=x,y,z}\sum_{\langle i, j\rangle_\alpha}  S^\alpha_i S^\alpha_j+
    g\sum_{[ijk]_{\alpha\beta\gamma}}S^\alpha_iS^\beta_jS^\gamma_k.\label{eq:Kitaev_spin}
\end{equation}
By using the Jordan-Wigner transformation~\cite{nasu2014vaporization}, one can transform the spins to Majorana fermions $c$ and $\bar{c}$, which obey $\{c_i,\bar{c}_j\}=0$ and  $\{c_i,c_j\}=2\delta_{ij}=\{\bar{c}_i,\bar{c}_j\}$. Then, the Hamiltonian becomes (see Refs.~\cite{nasu2014vaporization,nasu2017thermal} and Fig.~\ref{figA3})
\begin{eqnarray}
        H&=&-\frac{i}{4}\sum_{j\in {\rm filled\; circles}} J(c_jc_{j_x}+c_jc_{j_y}+\eta_b c_jc_{j_z})\nonumber\\
        &&+\frac{ig}{8}\sum_p(c_{p_1}c_{p_3}+\eta_{b_2}c_{p_3}c_{p_5}+\eta_{b_1}c_{p_5}c_{p_1}\nonumber\\
        &&\qquad\quad +c_{p_4}c_{p_6}+\eta_{b_1}c_{p_6}c_{p_2}+\eta_{b_2}c_{p_2}c_{p_4}).\nonumber \\
        &\equiv& \frac{i}{4}\sum_{jj^\prime}A_{jj^\prime}(\eta) c_j c_{j^\prime}\; . 
        \label{MajH}
\end{eqnarray}
Here, $\eta_b=i\bar{c}_j\bar{c}_{j_z}$, takes the value $\pm1$ on each $z$ bond $b=\langle jj_z\rangle_z$, $p$ denotes a single plaquette of the honeycomb lattice, $b_1$ and $b_2$ are the two $z$ bond on the plaquette, and $j$ denote the sites belonging to one sublattice (``filled circles"); see Fig.~\ref{figA3}. Compared to the 4-Majorana description of Ref.~\cite{kitaev2006anyons}, this description does not require projecting back to the physical Hilbert space, since the Hilbert space dimension remains the same in the spin and the Majorana representations.

For the Kitaev model, it is well known that, for each plaquette $p$ [see Fig.~\ref{figA3}], there exists a local conserved quantity, the plaquette operator~\cite{kitaev2006anyons,nasu2014vaporization}:
\begin{equation}
\label{vortex-spin}    W_p=\sigma_{p_1}^x\sigma_{p_2}^z\sigma_{p_3}^y\sigma_{p_4}^x\sigma_{p_5}^z\sigma_{p_6}^y.
\end{equation}
Here, $\sigma^\alpha_j = 2S^\alpha_j$ are the Pauli matrices.
The eigenvalues of $W_p$ are $\pm1$ and under the Jordan-Wigner transformation, the plaquette operator can be written as~\cite{nasu2014vaporization}
\begin{equation}
    W_p=\prod_{b\in p}\eta_b.  \label{vortex}
\end{equation}
By convention, states with $W_p=1$ on all plaquettes are defined to have a vortex- (or flux-)free configuration~\cite{kitaev2006anyons}. 

Since $[H,W_p]=0$, any state of the original many-body spin system can be written as a product state over the vortex ($W$) and Majorana ($M$) sectors. In particular, for the ground state, we have
\begin{equation}
    |G\rangle=|G_W\rangle |G_M\rangle.
    \label{eq:spin_GS}
\end{equation}
In the following, we discuss the two sectors separately.

\subsection{Vortex sector}

Let us start with the vortex sector. 
According to Lieb's theorem~\cite{lieb1994flux,kitaev2006anyons}, in the thermodynamic limit the $\eta$-dependent many-body ground state of the original spin system, is given by the vortex-free field configuration $|G_{W=+1}\rangle$. 
For the planar geometry used in this work, $\eta_b$ eigenvalues defined on the same row should have the same sign, which can be defined as $\eta_{r}$ with $r$ indicating the $r$th row (see Fig.~\ref{figA3}). Then, since there are $L_y-1$ rows of $\eta$ for a planar geometry with $L_y$ unit cells along the $y$-direction, the vortex-free configuration is actually highly degenerate, with the degeneracy being $2^{L_y-1}$. Nevertheless, these different $\eta_b$ configurations for a system with a vortex-free planar geometry are actually equal to each other because they can be transformed into the all $\eta_b=1$ configuration by the gauge transformation 
\begin{equation}
    \hat{c}_{j}=c_{j}\prod_{r<j_r} {\eta_r},
\end{equation}
with $j_r$ denoting the row index for site $j$; see Fig.~\ref{figA3}.

Similarly, for the vortex-free cylinder system, there are $2^{L_y}$ different configurations of $\eta_r$ that keep the system vortex-free. Nevertheless, they can be divided into two different configurations, defined by the even or odd number of rows with  $\eta_r=-1$. Moreover, in the case of all $\eta_b=1$, if, instead of a periodic boundary condition on the cylinder, we use an anti-periodic boundary condition, we will obtain a configuration with only one row of $\eta_b=-1$. This means that the two different $\eta_b$ configurations have the same topological properties in the thermodynamic limit. 

Considering that our probe does not alter the $\eta_r$ configuration and the states with different $\eta_r$ configurations are orthogonal, the detected signals from different $\eta_r$ configurations do not interfere with each other. Therefore, we can fix the gauge, and focus on the case where $\eta_b=1$.

\subsection{Majorana sector}

Replacing $\eta_b$ by their eigenvalues, a system with $2N$ sites gives rise to a $2N\times 2N$ skew-symmetric matrix $A_{jj'}=-A_{j'j}$, which defines a quadratic Hamiltonian describing the Majorana sector:
\begin{equation}
    H_M=\frac{i}{4}\sum_{jj^\prime}A_{jj^\prime}c_jc_{j^\prime}. 
    \label{anyM}
\end{equation}
The many-body Majorana ground state $|G_M\rangle$ can be constructed from the single-particle modes. For a $2N\times 2N$ single-particle Majorana Hamiltonian $H_M$, the creation operator for the $i$-th single-particle eigenmode can be written as 
\begin{equation}
    \label{eq:gamma}
    \gamma^\dagger_i=\sum_{j=1}^{2N} f^{i}_{j}c_j.
\end{equation}
Here, $f^i_j$ is the $j$th expansion coefficient of the $i$th eigenmode. Then, the ground state in the Majorana sector is given by~\cite{nasu2017thermal}
\begin{equation}
    |G_M\rangle=\gamma^\dagger_1\gamma^\dagger_2....\gamma_N^\dagger |0\rangle, \label{groundstate}
\end{equation} 
with the eigenvalues corresponding to the eigenmodes $\gamma_i^\dagger$ ($i=1,\dots,N$) being negative.

Assuming translation invariance (all $\eta_b=1)$, we can also derive the Majorana spectrum for a system with periodic boundary conditions in momentum space. 
To specify the unit cell notation in the Majorana representation, we represent the site index $j$ of the Majorana operator $c_j$, as the composite index $(m,l)$, with $l$ running over the two sites in the unit cell $m$.
Following Ref.~\cite{nasu2017thermal} let us introduce the (inverse) Fourier transform of $c_j=c_{m,l}$ as  
\begin{equation}
    c_{m,l}=\sqrt{\frac{2}{N}}\sum_{\bf q}
    \mathrm 
    e^{i{\bf q}\cdot{\bf r}_{m,l}}c_{{\bf q}l}.
\end{equation}
Here, the extra factor $2$ comes from the anti-commutation relation of Majorana fermions, $\{ c_{m,l}, c_{n,l^\prime}\}=2\delta_{mn}\delta_{ll^\prime}$. With this definition, the complex fermion $c_{{\bf q}l}$ fulfills the proper fermionic anticommutation relation
$\{c_{{\bf q}l},c^\dagger_{{\bf q}^\prime l^\prime}\}=\delta_{{\bf q},{\bf q}^\prime}\delta_{l,l^\prime}$,
and the Hamiltonian from Eq.~(\ref{MajH}) becomes
\begin{equation}
    H_M= \sum_{{\bf q}} \sum_{l,l^\prime} H^{l,l^\prime}_M({\bf q})\; c^\dagger_{{\bf q}l} c_{{\bf q}l^\prime}
\end{equation}
with 
\begin{equation}
    H^{l,l^\prime}_M({\bf q})=\frac{i}{2}\sum_{m,m'}  \mathrm  e^{-i{\bf q}\cdot{({\bf r}_{m,l}-{\bf r}_{m^\prime,l^\prime}})}A_{ml,m^\prime l^\prime}\; . 
    \label{eq:HmK}
\end{equation}
Thus, it follows that the spectral properties are determined by the eigenvalues of the matrix $H_M({\bf q})\sim iA/2$~\cite{nasu2017thermal}.

In the $\eta_b=1$ sector investigated in this work, if one simply assumes a periodic boundary condition for Eq.~(\ref{MajH}), the matrix $H_M({\bf q})$ in $(l,l^\prime)$-space can be shown to be 
\begin{widetext}
\begin{eqnarray}
    H_M({\bf q})&=&
    -i\frac{J}{4} \left(
\begin{array}{cc}
 0 & -2\cos(\frac{\sqrt{3}}{2}q_x)\mathrm e^{\frac{i}{2}q_y}-\mathrm e^{-i q_y} \\
 2\cos(\frac{\sqrt{3}}{2}q_x)\mathrm e^{-\frac{i}{2}q_y}+\mathrm e^{i q_y} & 0 \\
\end{array}
\right)\nonumber \\
&& +\frac{g}{4} \left(
\begin{array}{cc}
 \sin(\sqrt{3}q_x)-2\cos(\frac{3}{2}q_y)\sin(\frac{\sqrt{3}}{2}q_x) & 0 \\
 0 & -\sin(\sqrt{3}q_x)+2\cos(\frac{3}{2}q_y)\sin(\frac{\sqrt{3}}{2}q_x) \\
\end{array}
\right)\; .
\end{eqnarray}
Here, the distance between two nearby sites sets the unit length, and $q_x$ and $q_y$ are the two component of the quasimomentum vector ${\bf q}$. The dispersion relation reads as
\begin{eqnarray}
    \varepsilon_\pm({\bf q}) &=& 
    \pm\frac{1}{4} \Bigg(-\cos \sqrt{3} {q_x} \left(g^2 \cos 3 {q_y}+g^2-2 J^2\right)-2 g^2 \cos \frac{\sqrt{3} {q_x}}{2} \cos \frac{3 {q_y}}{2}+2 g^2 \cos \frac{3 \sqrt{3} q_x}{2} \cos \frac{3 {q_y}}{2}
    \nonumber\\
    && \qquad -\frac{1}{2} g^2 \cos 2 \sqrt{3} {q_x}+g^2 \cos 3 {q_y}+\frac{3 g^2}{2}+4 J^2 \cos \frac{\sqrt{3} {q_x}}{2} \cos \frac{3 {q_y}}{2}+3 J^2 \Bigg)^{1/2}\; .
\end{eqnarray}
\end{widetext}
From this equation, the Dirac points can be shown to be located at ${\bf q}=\left(\pm \frac{4\pi}{3\sqrt{3}},0\right)$, with the energy gap $\Delta=3\sqrt{3}g/4$.

Hence, experiments on finite-size systems, aiming to detect the topological properties of the Majorana bands, should select the underlying lattice geometry carefully.
For instance, to hit the Dirac point (where the Berry curvature of the Majorana bands is strongest) in small systems, the number of the unit cells in each direction should be a multiple of $3$.

\section{\label{app:factor_2}Time evolution in the Majorana sector, and the origin of the conversion factor of 2 occurring in the dynamical probes}

In this section, we recall the single-particle formalism we use to treat the Majorana sector of the model. We also explain the origin of the additional conversion factor of $2$ between the dynamically extracted data for ($\Delta$, $v_\mathrm{edge}$) in the main text, and their actual (theory) values.

To time-evolve the system, consider again a generic quadratic Hamiltonian $H_M$ in the Majorana representation,
\begin{equation}
    H_M(A)=\frac{i}{4}\sum_{jj^\prime}A_{jj^\prime}c_jc_{j^\prime}. 
    %\label{anyM}
\end{equation}
Using the normalization factor $1/4$, this operator fulfills the relation~\cite{kitaev2006anyons} 
\begin{equation}
    [-iH_M(A),-iH_M(B)]=-iH_M([A,B]), \label{commutM}
\end{equation}
where the commutator on the left-hand-side acts on Fock space, while the commutator on the right-hand-side -- on the $2N\times2N$ dimensional space associated with the skew-symmetric matrix $A$.
Then, the time evolution of Majorana modes in the Heisenberg picture can be calculated as
\begin{equation}
    \mathrm e^{itH_\alpha(A)}c_{j}\mathrm e^{-itH_\alpha(A)}=\sum_{j^\prime} c_{j^\prime}Q_{j^\prime j}, 
    \label{evoluM}
\end{equation}
with $Q=\mathrm e^{-tA}$. 

We are now in a position to explain the origin of the conversion factor of $2$ between the dynamically extracted gap and edge velocity, and their actual theoretical values. For our Majorana system, it follows from the relation $Q=\mathrm e^{-At}$ that the dynamics is described by the eigenvalues of the matrix $iA$ [instead of $iA/4$, as one may naively think due to Eq.~\eqref{anyM}]. This factor is important since it scales time in Eq.~\eqref{evoluM}.
Let us now contrast this with the static properties of the Majorana system. For instance, to get the many-body gap, we use Eq.~\eqref{eq:HmK}, which states that the gap is determined by the eigenvalues of the matrix $iA/2$~\cite{nasu2017thermal}.
Comparing the dynamics generated by $iA$, to the spectral properties obtained from $iA/2$, we find the additional factor of $2$, required for the conversion between the two. 
The physical origin of this factor is intimately related to the fractional character of the Majorana quasiparticles: it can be traced back to the restriction that excitations can only be created in pairs in the Kitaev model~\cite{kitaev2006anyons}.
Finally, let us mention that we also confirmed the existence of this factor in the full spin system, Eq.~\eqref{eq:H_eff}, by extracting the many-body gap from spectroscopy.

\section{\label{app:Floquet_Majorana}Floquet drive in the Majorana sector}

In the presence of the Floquet drive, the time evolution operator over one Floquet period reads as
\begin{equation}
    U_F=\mathrm e^{-iTH_x(J_x)}\mathrm e^{-iTH_y(J_y)}\mathrm e^{-iTH_z(J_z)},
\end{equation}
where $H_\alpha=- \sum_{\langle j,j'^\prime \rangle_\alpha}
J_\alpha S^\alpha_j S^\alpha_{j^\prime}$.
Note that the plaquette operators $W_p$ [Eq.~(\ref{vortex})] commute with all $H_\alpha$, which means they also commute with $U_F$. Hence, the Floquet drive does not mix the vortex and Majorana sectors, and we can numerically simulate the dynamics of the spin system in the Majorana sector.

Hence, we can restrict the analysis to the dispersive Majorana sector: 
\begin{equation}
H_\alpha(J_\alpha)=\frac{i}{4}\sum_{\langle j,j^\prime \rangle_\alpha} A^\alpha_{jj^\prime}(J_\alpha) c_j c_{j^\prime},
\end{equation}
which is obtained from the original spin Hamiltonian $H_\alpha=- \sum_{\langle j,j^\prime \rangle_\alpha}
J_\alpha S^\alpha_jS^\alpha_{j^\prime}$ using the Jordan-Wigner transformation, cf.~Sec.~\ref{app:mapping}.

Following Eq.~({\ref{evoluM}}), the time evolution of a Majorana operator becomes
\begin{eqnarray}
    U_F^\dagger c_j U_F&=&\sum_{j^\prime}c_{j^\prime}[\mathrm e^{-TA^z}\mathrm e^{-TA^y}\mathrm e^{-TA^x}]_{j^\prime j}\nonumber \\
    &\equiv&\sum_{j^\prime} c_{j^\prime} [\mathrm e^{-TA^\mathrm{eff}}]_{j^\prime j}\; . \label{eq:Aeff}
\end{eqnarray}  
Note that the order of unitaries in the definition of $A_\mathrm{eff}$ is reversed compared to $U_F$. With this definition of $A_\mathrm{eff}$, the Floquet Hamiltonian is given by
\begin{equation}
    H_\mathrm{eff}=\frac{i}{T}\log U_\mathrm{F} \equiv\frac{i}{4}\sum_{k,l}A^\mathrm{eff}_{kl}c_kc_l.\label{eq:Heff_Floq}
\end{equation}
There is no handy closed-form expression for the exact matrix $A^\mathrm{eff}$. However, for sufficiently high drive frequencies, an approximation can be constructed using the inverse-frequency expansion.

\section{\label{app:spetroscopy}Details on the protocol for gap spectroscopy}

Here we discuss in detail the spectroscopy protocol in the presence of the Floquet drive. We then go on and show supplementary results to the main text.  

We prepare the system in the Majorana ground state of the isotropic effective Hamiltonian $H_\mathrm{eff}$, whose eigenmodes are annihilated by the operators $\gamma_i$.
We then evolve the system in the presence of the Floquet drive and simultaneously apply the spectroscopy protocol; this gives rise to the modified Floquet unitary
\begin{equation}
    U_F[\ell]=\mathrm e^{-iTH_x(J)}\mathrm e^{-iTH_y(J)}\mathrm e^{-iTH_z(J_z(\ell))},
\end{equation}
where the $z$-bond strength changes at a frequency $\omega_p\ll\omega_D$, according to:
\begin{equation}
    J_z(\ell)=J[1+A_p\sin{(\omega_p\ell T)}],
\end{equation}
with $A_p$ the (dimensionless) probe strength, and $\ell$ -- the Floquet cycle. Treating the spectroscopic probe as a constant during the cycle is justified by the scaling of the topological gap we want to detect: $\omega_p\sim \Delta\propto\omega_D^{-1}$. 

Thus, at times $t_n=n T$ ($n\in\mathbb{N}$), the time-evolution operator is given by
\begin{equation}
    U_F(t_n) =\prod_{\ell=0}^{n} U_F[\ell]\; .
\end{equation}

\begin{figure}[t!]
\includegraphics[width = \linewidth]{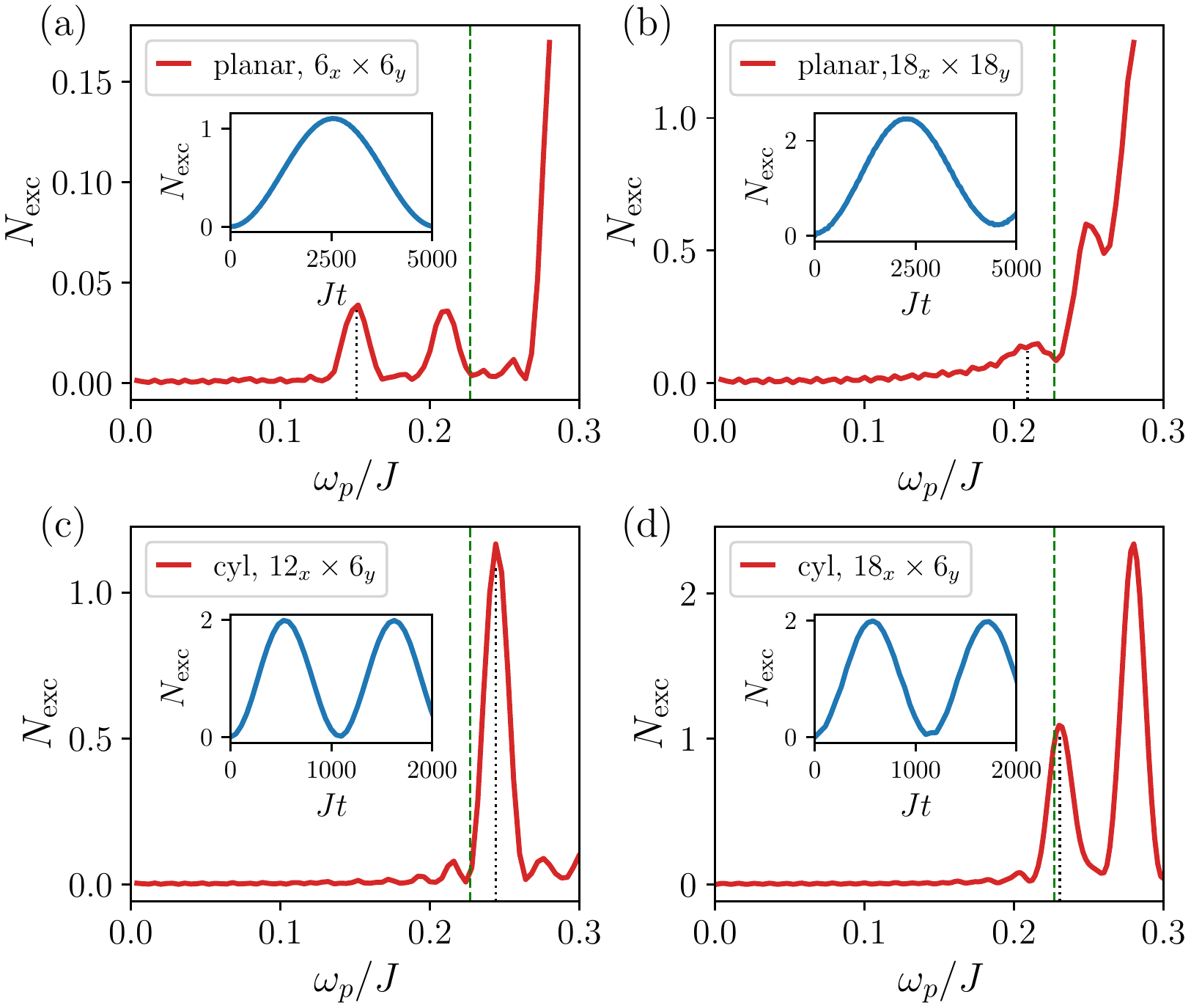}
\caption{The number of excited particles depends on the probe frequency $\omega_p$, for both planar and cylinder geometry with different system sizes (see legend). 
{\bf (a)} For the planar geometry with a small system size of $6_x\times6_y$, the boundary states have a non-negligible influence which shows up as an excitation of a mid-gap state. 
{\bf (b)} For a large system size, $18_x\times18_y$, the influence of the boundary states becomes negligible and the lowest resonant peak indicates the location of the bulk gap. 
{\bf (c)} and {\bf (d)} For a cylindrical geometry, the lowest resonant peak indicates the location of the bulk gap quite well for both $12_x\times 6_y$ and $18_x\times6_y$ systems. 
The black dotted vertical line indicates the lowest resonant frequency $\omega_\mathrm{res}$, and the green dashed line indicates the location of $2\Delta$. The insets show the time dependence of the signal $N_\mathrm{exc}(t)$ at $\omega_p=\omega_\mathrm{res}$ used to extract the bulk gap. The other parameters are the same as in Fig.~\ref{fig_2}, i.e., $Jt_\ast=300$, $A_p=0.02$, $\omega_D/J=12$. }
\label{figA4}
\end{figure}

The spectroscopic signal is proportional to the number of excitation particles $N_\mathrm{exc}$, defined as
\begin{equation}
    N_\mathrm{exc}(t_n)=N-\langle G_M|U^\dagger_F(t_n)\sum_{i=1}^{N}\gamma^\dagger_i\gamma_i U_F(t_n)|G_M\rangle,
\end{equation}
where $N$ is the number of occupied Majorana modes in the initial state (out of $2N$ modes in total), and $\gamma_i$ is defined in terms of the real-space Majorana operators in Eq.~\eqref{eq:gamma}. A straightforward application of Eq.~\eqref{evoluM} leads to the expression 
\begin{equation}
    N_\mathrm{exc}(t_n)=N-\sum_i F^T(t_n) C F^\ast(t_n), \label{eq:Nexc}
\end{equation}
where the $2N\times 2N$ correlation matrix $C$ has the matrix elements $C_{ij}=\langle G_M|c_ic_j|G_M\rangle$, and
\begin{equation}
    F(t_n){=}\prod_{\ell=1}^{n} 
\mathrm e^{-{\ell T}A^z(J_z(\ell))}
\mathrm e^{-{\ell T}A^y(J)}
\mathrm e^{-{\ell T}A^x(J)} f(0)
\end{equation}
is an $N$-dimensional vector of time-evolved expansion coefficients, cf.~Eq.~\eqref{eq:gamma}.

Using Eq.~(\ref{eq:Nexc}), we calculate the dependence of $N_\mathrm{exc}$ on the probe frequency $\omega_p$ for different geometries, cf.~Fig.~\ref{figA4}. 
For the planar geometry, we see a pronounced finite-size effect in small systems: for the $6_x\times6_y$ system shown in Fig.~\ref{figA4}(a), the lowest resonant peak in the spectroscopy signal falls inside the gap due to the presence of boundary state excitations. Yet, for a large planar system, e.g., $18_x\times18_y$, the influence of the boundary states becomes weaker, and the lowest resonance peak shown in Fig.~\ref{figA4}(b) gets close to $2\Delta$ [see  Fig.~\ref{fig_2}(d)].  
On the other hand, the influence of the boundary states is less pronounced in the cylinder geometry. For the $12_x\times6_y$ and $18_x\times6_y$ systems shown in Fig.~\ref{figA4}(c) and (d), both of their first resonant peaks are close to $2\Delta$. Due to finite-size effect, the resonant peak in the larger system is closer to the theory-predicted gap value, when compared to the smaller system, especially for a system with a small gap [see also Fig.~\ref{fig_2}(d)].

\begin{figure}[t!]
\includegraphics[width = 0.67\linewidth]{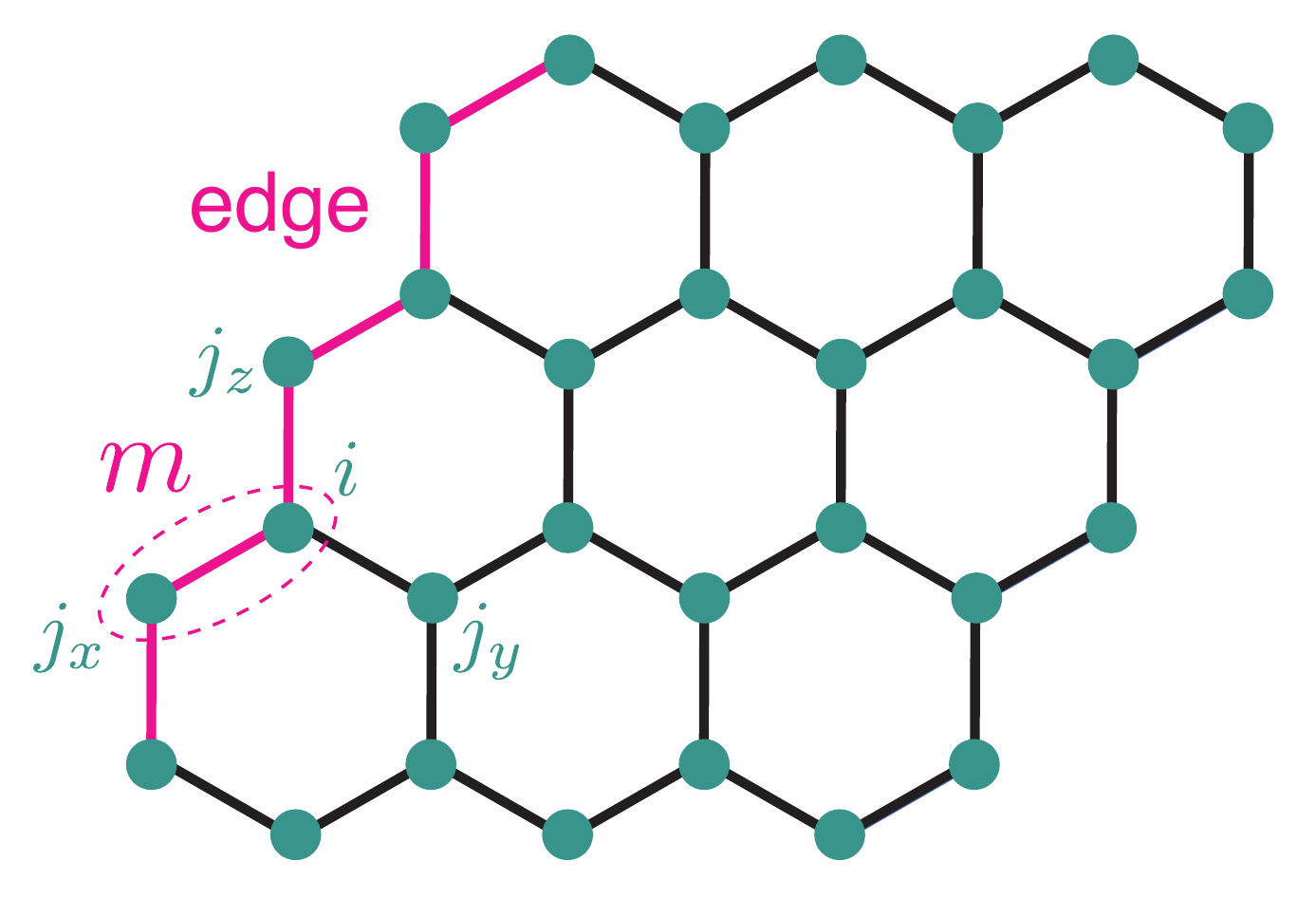}
\caption{
Schematic representation of the notation for the quench protocol: $m$ labels a unit cell at the edge of the system; the site $i$ belongs to the unit cell $m$, while $\langle i,j_x\rangle_x$, $\langle i,j_y\rangle_y$, and $\langle i,j_z\rangle_z$ indicate the bonds connected to $i$ that are considered in the expression for the local unit cell energy $H_m$, cf.~Eq.~\eqref{eq:quench}. 
}
\label{fig_def_edge}
\end{figure}

\section{\label{app:chiral_transport}Details on the protocol for chiral edge transport}

In the main text, we showed that the chiral property of $H_\mathrm{eff}$ can be detected by measuring the time-evolution of the $zz$-correlations on the edge, following a local quench. We motivated this measurement by an analogy with edge transport of energy. In this section, we show numerical results on the chiral energy transport.

In the Majorana sector, the Hamiltonian can be written as a sum over unit cells labelled by $m$ and $n$:
\begin{equation}
    H_M=\sum_{ml,nl^\prime} H_{M;ml,nl^\prime},
\end{equation}
where $l$ and $l^\prime$ mark the lattice sites in unit cell $m$ and $n$. 
Let us define the energy operator in a local unit cell as
\begin{eqnarray}
    H_m &=&\frac{1}{2}\sum_{l,nl^\prime} H_{ M;ml,nl^\prime}+H^\dagger_{M;ml,nl^\prime},\nonumber\\
    &=& \frac{J}{2} \sum_{i\in m} S^x_iS^x_{j_x} + S^y_iS^y_{j_y} + S^z_iS^z_{j_z}\; ,
\end{eqnarray}
see Fig.~\ref{fig_def_edge} for details of the notation.
The motivation behind this definition goes as follows. Let $|\phi\rangle$ be an eigenvector of $H_M$ with the corresponding eigenvalue $E_\phi$, and let $|\phi\rangle_{nl^\prime}$ denote the corresponding components of $|\phi\rangle$ on the $l^\prime$-th site of the $n$-th unit cell: $\sum_{nl^\prime}H_{M;ml,nl^\prime}|\phi\rangle_{nl^\prime}=E_\phi|\phi\rangle_{ml}$. Hence, the expectation value of $H_m$ in this state is 
\begin{eqnarray}
\langle \phi|H_m|\phi\rangle =  \langle \phi|\sum_l E_\phi|\phi\rangle_{ml}= E_\phi \sum_l \langle\phi|\phi\rangle_{ml}.
\end{eqnarray}
From this equation, one finds that the expectation is just the number of particles (i.e., the occupation) in the $m$-th unit cell  times the corresponding energy, which is a good description of the energy in the $m$-th unit cell.

\begin{figure}[t!]
\includegraphics[width = \linewidth]{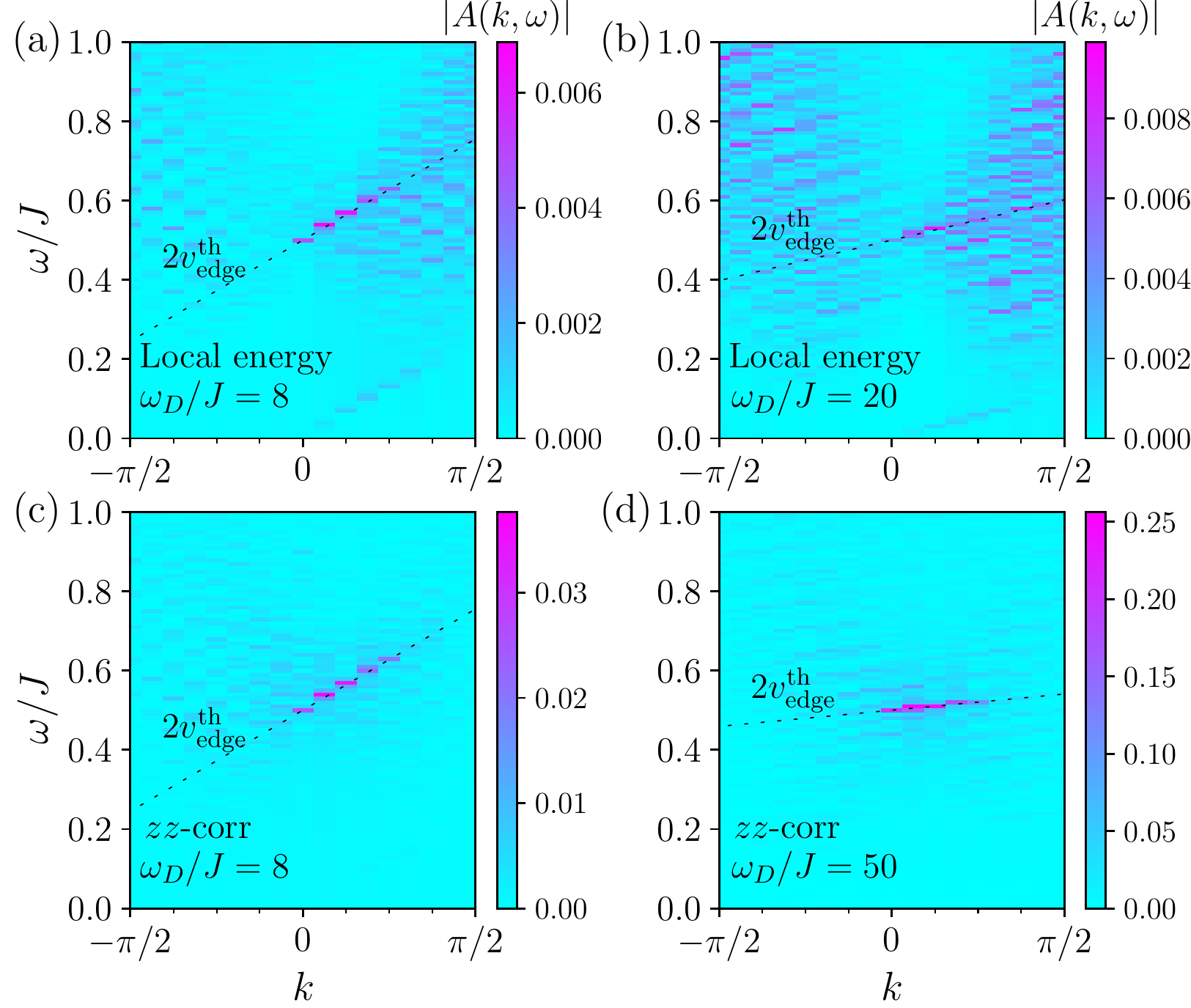}
\caption{
Fourier spectrum $|A(k,\omega)|$ of the local unit cell energy and the $zz$-correlations following the boundary quench in a $16_x\times32_y$ system.  
{\bf (a)} and {\bf (b)} Fourier spectrum for the local unit cell energy for $\omega_D/J=8, 20$, and $N_T=801,2001$, respectively. The chiral signal is clearly noticeable among additional excitations. 
{\bf (c)} and {\bf (d)} Fourier spectrum for the $zz$-correlations, for $\omega_D/J=8,50$, and $N_T=801,5001$, respectively. The dominant chiral behavior is clearly discernible, and the amount of non-chiral excitations appears reduced. 
For all panels, the slope of the black dashed line corresponds to the theory prediction, $2v^\mathrm{th}_\mathrm{edge}{\times} k{+}0.5$, and $J_q/J=0.008$. For comparison, in Fig.~\ref{fig_3} of the main text, the black dashed line corresponds to a fit [cf.~Sec.~\ref{app:fitting_procedure}].
}
\label{figA5}
\end{figure}
	
To get the chiral edge-state transport, we take a system with a boundary, and prepare it in the ground state of $H_\mathrm{eff}$, cf.~Eq.~\eqref{groundstate}. Then we apply a perturbation $H_p$ as a quench to the effective Hamiltonian, acting in the middle of the edge:
\begin{equation}
    H_p=J_q S^z_iS^z_j,
\end{equation}
with $J_q$ being the perturbation strength, and $\langle ij\rangle_z$ denotes the $z$-bond in the middle of the edge. 
The corresponding Floquet unitary after the quench thus becomes
\begin{equation}
    \tilde U_F=\mathrm e^{-iTH_x(J)}\mathrm e^{-iTH_y(J)}\mathrm e^{-iT\left(H_z(J){+}H_p(J_q)\right)},
\end{equation}
Finally, we evolve the system with $\tilde U_F$, and measure
the local unit cell energy $H_m$ at the $m$-th unit cell at the edge: 
\begin{equation}
    E_m(\ell)=\langle G_M| \left[\tilde U_F^\dagger\right]^\ell H_m  \left[\tilde U_F\right]^\ell|G_M\rangle,
\end{equation}
which can be evaluated in practice with the help of Eq.~\eqref{evoluM}. 

Similarly, in Sec.~\ref{subsec:edge_transport} of the main text we defined the dynamics of the local $zz$-correlator, $C_m(\ell)$, after the quench.

The Fourier spectrum of $E_m(t)$ can be obtained from 
\begin{equation}
    \label{eq:FourierTrans}
    A(k^j_y,\omega_l)=\sum^{L_y}_{m=1}\sum_{\ell=1}^{N_T} E_m(\ell)\; \mathrm e^{-ik^j_y m +i \omega_l \ell T},
\end{equation}
where, $L_y$ is the number of unit cells along the $y$ direction, $N_T$ is the number of Floquet cycles used to do the Fourier transformation, $k^j_y=2\pi j/L_y$, and $\omega_\ell=2\pi \ell/(TN_T)$.
In the main text, to simplify notation, we used the quantities $k$ and $\omega$.
We note that since $E_m(\ell)$ is real-valued, we have the reflection symmetry $|A(k,\omega)|=|A(-k,-\omega)|$.

In our calculation of $H_m$, considering that the strength $g$ of the three-spin correction term is much smaller than $J$, we drop all terms proportional to $g$ in Eq.~(\ref{MajH}) to simplify the calculation. We show the obtained Fourier spectrum for the local unit cell energy in Fig.~\ref{figA5} (a) and (b), corresponding to drive frequency $\omega_D/J=8,20$, respectively. The data show a strong signal due to the excitation of the chiral edge modes, which agrees well with twice the theoretical group velocity $v^\mathrm{th}_\mathrm{edge}=3\Delta/(2\pi)$, as shown by the thin black dashed line in the figures. Note that in Sec.~\ref{subsec:edge_transport}, we only showed results for the $zz$-correlation [Eq.~(\ref{eq:quench_C})], instead of the local unit cell energy for the purpose of simplifying the experimental realization. Here, for completeness, we also show data for the $zz$-correlation in Fig.~\ref{figA5} (c) and (d) [with $\omega_D/J=8,50$, respectively]. Comparing Fig.~\ref{figA5}(a) and (c), one finds that the spectrum for the $zz$-correlation may not show all excitations, but it captures the major chiral signal of the local unit cell energy results; moreover, the signal is cleaner for the $zz$-correlations, which is advantageous in determining the value of $v_\mathrm{edge}$ using a linear fit [see Sec.~\ref{app:fitting_procedure}]. Especially when the topological gap is small, e.g., in Fig.~\ref{figA5}(d) where $\omega_D/J=50$, the agreement of the chiral signal with the theoretical prediction for $v^\mathrm{th}_\mathrm{edge}$ is particularly clear. However, the corresponding signal for the local unit cell energy is surrounded by additional excitations as shown by the Fig.~\ref{figA5}(b).

\begin{figure}[t!]
\includegraphics[width = \linewidth]{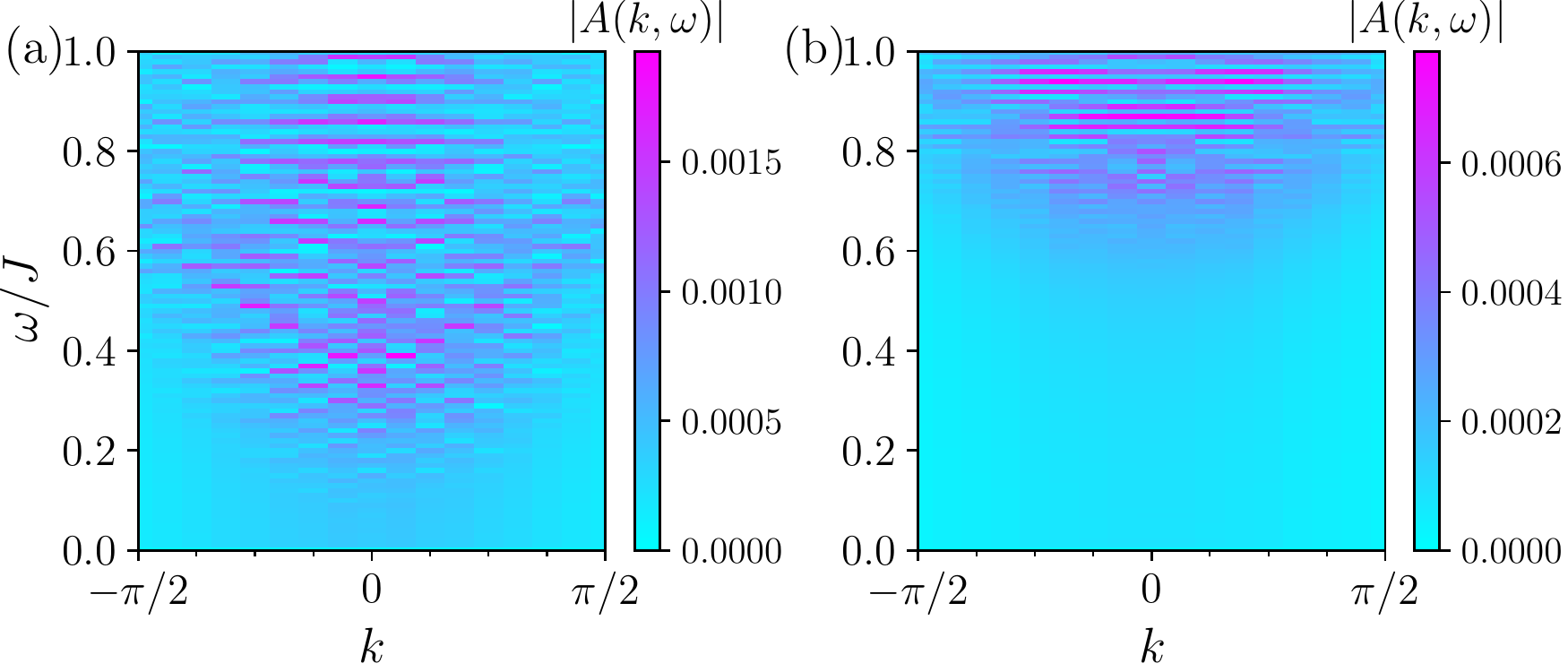}
\caption{
Fourier spectrum $|A(k,\omega)|$ of the $zz$-correlations at the boundary of a $16_x\times32_y$ system following the quench [see text], for two different points belonging to phase $A_z$ of the phase diagram [Fig.~\ref{fig_1}c]:  
{\bf (a)} At the critical point ($J_x{=}J_y{=}0.5J$ and $J_z{=}J$) the chiral bulk gap in the Majorana spectrum closes [see Fig.~\ref{fig_4}(a)], and the chiral signal is lost.
{\bf (b)} The chiral signal is not observable deeper in the toric code phase ($J_x{=}J_y{=}0.25J$ and $J_z{=}J$) either. 
The figure demonstrates that the chiral signal in the Fourier spectrum is an intrinsic property of the chiral spin liquid phase $B$; in particular, it is not induced by the ``polarization'' of the Floquet drive.
The parameters are $\omega_D/J=8$, $J_q/J$=0.008, and $N_T=801$, and the remaining parameters are the same as in Fig.~\ref{figA5}(c).
}
\label{figF2}
\end{figure}

To further demonstrate that the chiral signal arises from the properties of the chiral spin liquid phase $B$, instead of being a mere consequence of the circular ``polarization'' of the Floquet drive, we display the Fourier spectrum of the boundary $zz$-correlation at the phase transition point, i.e., $J_x=0.5J=J_y$, and $J_z=J$ in Fig.~\ref{figF2}(a), as well as for one point in the bulk of the non-chiral toric code phase $A_z$ ($J_x=0.25J=J_y$, and $J_z=1$), see Fig.~\ref{figF2}(b). As expected, no chiral signal is observable outside the chiral spin liquid phase.

\begin{figure}[t!]
\includegraphics[width = 0.95\linewidth]{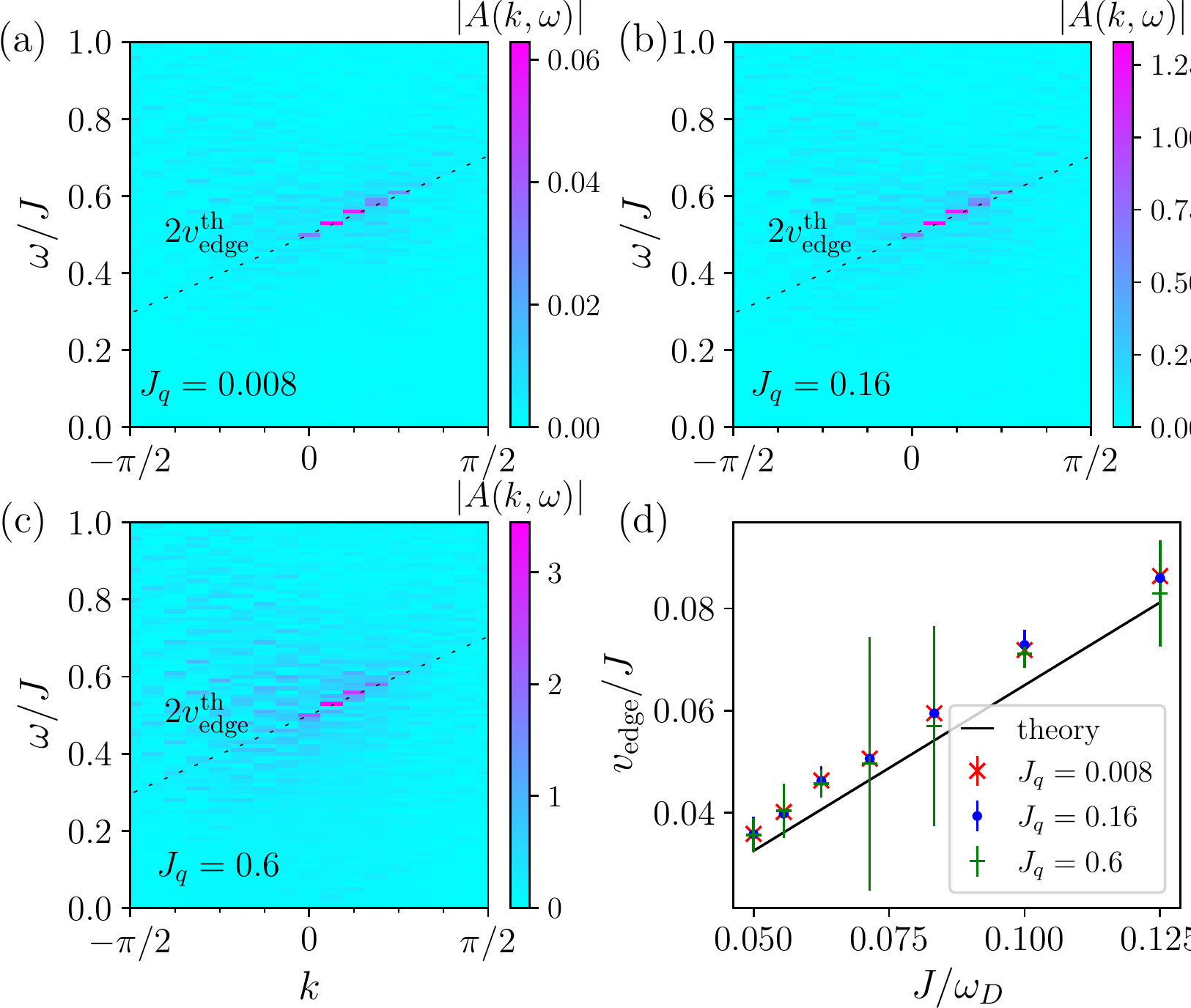}
\caption{The effect of increasing quench strength $J_q$ on the Fourier spectrum of the $zz$-correlation and corresponding edge state velocity. The plots show the Fourier spectrum of the $zz$-correlation for three different quench strengths: (a) $J_q/J=0.008$, (b) $J_q/J=0.16$, and (c) $J_q/J=0.6$. The corresponding extracted edge state velocity $v_{\rm edge}$ is shown in (d). As the quench strength increases, residual, ``noisy" components appear in the Fourier spectrum, but the dominant chiral signal remains qualitatively unchanged. The system size is $16_x\times32_y$ for a cylinder geometry. The model parameters used in the study are $\omega_D/J=10$, and $N_T =1001$, and the cutoff used in the fitting is 0.4. }
\label{figA8}
\end{figure}

Finally, we point out that we use a relatively small value $J_q/J=0.008$ in the numerical simulations in order to avoid unwanted additional excitations that may interfere with the process of extracting the edge-state velocity. With respect to the feasibility of a potential experimental realization, it is important to note that this value can be increased. Let us briefly examine the role of the quench strength $J_q/J$. The obtained results, shown in Fig.~\ref{figA8}, reveal that increasing $J_q/J$ introduces additional ``noise" to the Fourier spectrum, but the dominant chiral signal and the extracted edge-state velocity $v_{\rm edge}$ remain qualitatively unchanged even for values up to $J_q/J
\lesssim 0.6$.

\section{\label{app:ramp}Details on the protocol for ramping through the critical point}

We now turn our attention to the ramping process used to prepare the initial state for small-enough systems.  

To this end, we initialize the system in the ground state of $H_\mathrm{eff}$ at $J_x{=}0$, $J_y{=}0$, and $J_z{=}J$. To avoid problems with the massive ground state degeneracy, we pick the linear combination consistent with a vortex-free configuration; this is automatically the case when working in the Majorana representation at $\eta=1$.
We then, slowly ramp up the $x$ and $y$-couplings until we reach the isotropic point in the bulk of the chiral spin liquid phase at time $t=t_\mathrm{ramp}$. This amounts to the following protocol:
\begin{eqnarray}
    H_{x}(t)&=&-J_x\frac{t}{t_\mathrm{ramp}}\sum_{\langle jk\rangle_x}S^x_jS^x_k \nonumber\\
    H_{y}(t)&=&-J_y\frac{t}{t_\mathrm{ramp}}\sum_{\langle jk\rangle_y}S^y_jS^y_k\nonumber\\
    H_{z}&=&-J_z\sum_{\langle jk\rangle_z}S^z_jS^z_k.
\end{eqnarray}
If the ramp time is much longer than the Floquet period, $T\ll t_\mathrm{ramp}$, we can assume that the ramp is effectively constant within each drive cycle. Then the time evolution operator at drive cycle $\ell$ reads as
\begin{equation}
    U_F[\ell]=\mathrm e^{-iTH_x(\ell)}\;
    \mathrm e^{-iTH_y(\ell)}  \;
    \mathrm e^{-iTH_z} \; ,
\end{equation}
Following once again Eq.~\eqref{evoluM}, we can numerically compute the initial state after the ramp; we then use this state as the initial state for the quench protocol, following which we calculate the time evolution of the local unit cell energy, cf.~Sec.~\ref{app:chiral_transport}.

\begin{figure}[t!]
\includegraphics[width = \linewidth]{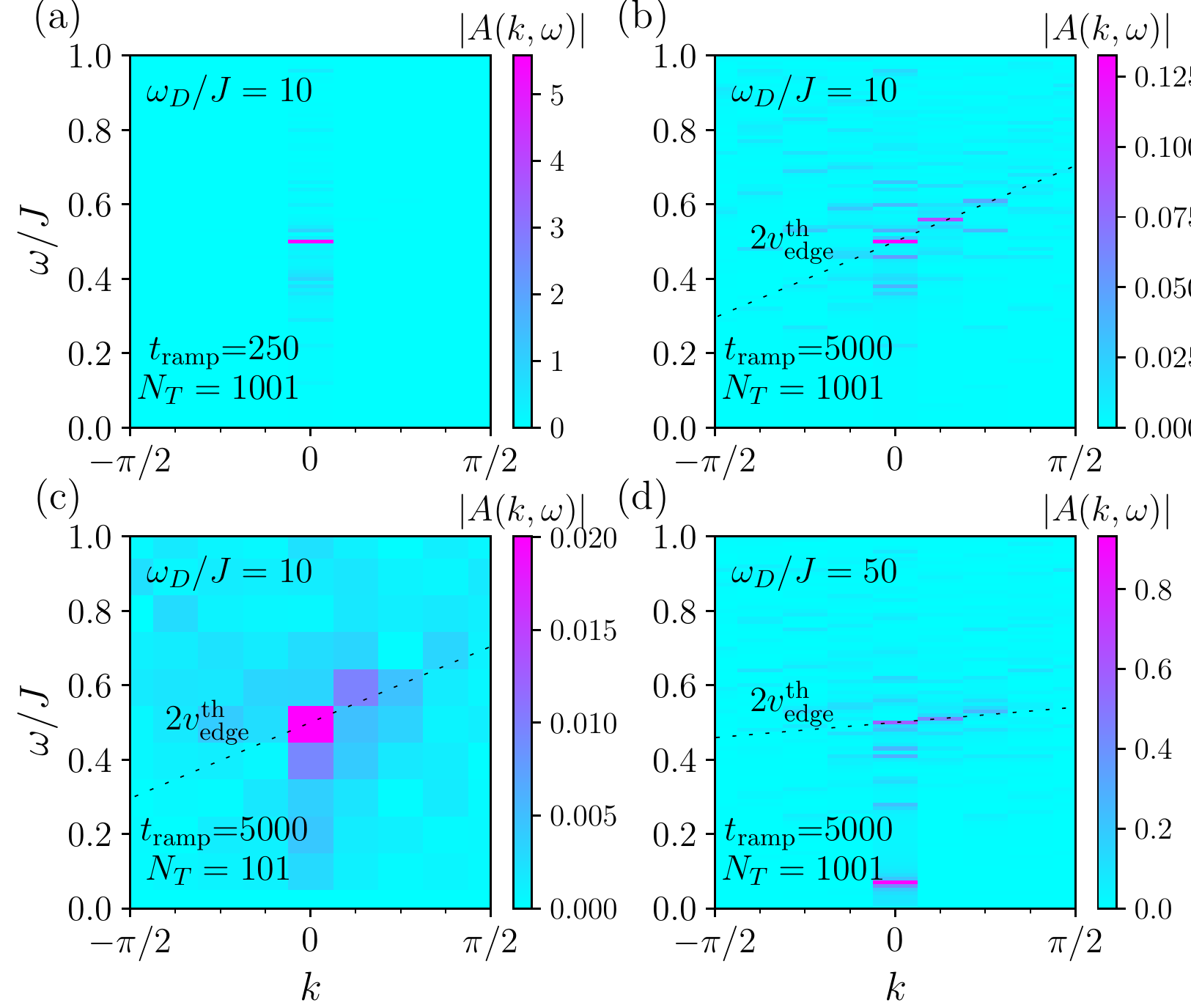}
\caption{
Fourier spectrum of the $zz$-correlation for various ramp and observation times.
{\bf (a)} For a short $Jt_\mathrm{ramp}=250$, strong signal appears at $k=0$, which dominates the chiral signal shown in Fig.~\ref{fig_4}(b). 
{\bf (b)} At long ramp durations, $Jt_\mathrm{ramp}=5000$, the dominance of the chiral signal is restored. For panels (a) and (b), $N_T=1001$ produces a good resolution of the signal. 
{\bf (c)} For short observation times, $N_T=101$, the resolution becomes low but the chiral behavior still exists. 
The drive frequency is $\omega_D/J=10$ for panels (a), (b), and (c). 
{\bf (d)} At large drive frequencies, e.g., $\omega_D/J=50$, a strong signal at $k=0$ comparable to the chiral signal persists even at long ramp times $Jt_\mathrm{ramp}=5000$.
The system size is $8_x\times16_y$ for all panels, and the slope of the black
dashed line corresponds to the theory prediction: $2v^\mathrm{th}_\mathrm{edge}$.
}
\label{figA6}
\end{figure}

The careful reader may have noticed that, in Fig.~\ref{fig_4}(b) of the main text, we did not display data in the region $k=0$ [see the greyed out region in Fig.~\ref{fig_4}(b)]: this is because for $Jt_\mathrm{ramp}=250$ used in Fig.~\ref{fig_4}(b), the Fourier signal at $k=0$ is too strong and dominates the chiral signal, as we show here in Fig.~\ref{figA6}(a). However, if the ramp time is long enough (compared to the inverse gap), the chiral signal eventually becomes dominant as shown in Fig.~\ref{figA6}(b), which is calculated with $Jt_\mathrm{ramp}=5000$.

In addition to the ramp time, the observation time after the quench, $t_\mathrm{obs}=T N_T$, also affects the Fourier spectrum. In particular, it influences the resolution along the $\omega$-axis via the Fourier transform, cf.~Eq.~(\ref{eq:FourierTrans}). In Fig.~\ref{figA6}(c), we show results for a smaller value of $N_T$. Among others, these data demonstrates that although a poor $\omega$-resolution may affect the slope $v_\mathrm{edge}$ [dashed black line], the signal still exhibits clean chiral properties.  

It is noted that for the small gap case with $\omega_D/J=50$ shown in Fig.~\ref{figA6}(d), there are still strong signals comparable to the chiral signal at $k=0$. Hence, to improve the accuracy of extracting the edge current velocity, we ignore the $k{=}0$ point.

\section{\label{app:lin_resp}Linear response analysis of the $zz$-correlations}

In order to understand the edge-state physics displayed by the $zz$-correlations after the quench, we apply linear response theory. The system we consider is an infinite cylinder, with all $\eta=1$. Then, the $zz$-correlation from Eq.~(\ref{eq:quench_C}) can be written as
\begin{equation}
    C_m(t)=\frac{iJ}{4}{\rm Tr}\left[\frac{1}{2}(c_{m,2}c_{m+1,1}-c_{m+1,1}c_{m,2})\rho(t)\right],
\end{equation}
where the notation is the same as in Fig.~\ref{fig_def_edge}, and $\rho(t)$ denotes the density matrix at time $t$. At $t=0$, the latter can be obtained from the ground state Eq.~(\ref{groundstate}). 

Using the Fourier transform
\begin{equation}
    c_{m,l}=\sqrt{\frac{2}{ N_y}}\sum_m e^{iqr_{m}}c_{q,l},
\end{equation}
the $zz$-correlation becomes
\begin{eqnarray}
    C(k,t)&=&\frac{iJ}{4}\sum_q{\rm Tr} [(c^\dagger_{q,2}c_{k+q,1}e^{i(k+q)\delta_z}\nonumber \\ && -c^\dagger_{q,1}c_{k+q,2}e^{-iq\delta_z})\rho(t)].
\end{eqnarray}
Here, $N_y$ is the number of the sites in the periodic direction, and $\delta_z=r_{m+1}-r_{m}$ is the distance between two nearby $z$-bonds at the edge.

In the interaction representation, $\rho_I(t)=e^{iH_0t}\rho(0)e^{-iH_0t}$, we have
\begin{equation}
\rho_I(t)=\rho(0)-i\int_0^t d\tau[H^\prime_I(\tau),\rho_I(\tau)],
\end{equation}
where, $H_0=H_{\rm eff}$ is given in Eq.~\eqref{eq:Heff_Floq}, and $H^\prime$ is the perturbation Hamiltonian with interaction picture representation $H^\prime_I(t)=e^{iH_0t}H^\prime e^{-iH_0t}$. 

Then, defining 
\begin{equation}
i (c^\dagger_{q,2}c_{k+q,1}e^{i(k+q)\delta_z}-c^\dagger_{q,1}c_{k+q,2}e^{-iq\delta_z})=c^\dagger_{q}c_{k+q},
\end{equation} 
and using the linear response approximation, i.e., replacing $\rho_I(\tau)$ by $\rho_I(0)$, one obtains
\begin{eqnarray}
    C(k,t)=-i\frac{J}{4}\sum_q \int^t_0 d\tau{\rm Tr} \{[e^{iH_0t}c^\dagger_{q}c_{k+q}e^{-iH_0t},H^\prime_I(\tau)]\rho_0\}.\nonumber\\ \label{eq:Ck_linear}
\end{eqnarray}
In doing so we neglected the term ${\rm Tr}[e^{iH_0t} c_{q,2}^\dagger c_{k+q,1}e^{-iH_0t}\rho_0]$ since it vanishes for $q\ne0$. 

We note that for a cylinder geometry with an even number $2N_x$ of sites in one unit cell, the Hamiltonian has exactly $N_x$ positive eigenvalues and $N_x$ negative eigenvalues for each momentum $q$. To make this explicit in the notation of the eigenstate $\gamma_{q,j}$, we define the index $j$ to range from $-N_x$ to $-1$ and from $1$ to $N_x$. Thus, the Hamiltonian can be written as 
\begin{equation}
H_0=\sum_{q,j}\epsilon_{qj}\gamma^\dagger_{q,j}\gamma_{q,j},\label{eq:H0_eigenform}
\end{equation}
with $\epsilon_{qj}$ the corresponding eigenvalues.
Moreover, due to the property of Majorana fermions, $\gamma^\dagger_{q,j}=\gamma_{-q,-j}$, we have the relation $\epsilon_{q,j}=-\epsilon_{-q,-j}$. Therefore, $\{\gamma_{qj},\gamma^\dagger_{q^\prime j^\prime}\}=\delta_{qq^\prime}\delta_{jj^\prime}$ and $\{\gamma_{qj},\gamma_{q^\prime j^\prime}\}=\delta_{-qq^\prime}\delta_{-jj^\prime}$. 

With these definitions, the perturbation term can be written as 
\begin{equation}
    H^\prime=\sum_{q,q^\prime,j,j^\prime} V^{j,j^\prime}_{q,q^\prime} \gamma^\dagger_{qj}\gamma_{q^\prime j^\prime},
\end{equation}
and one can show that 
\begin{equation}
V^{-j^\prime,-j}_{-q^\prime,-q}=-V^{j,j^\prime}_{q,q^\prime}.
\end{equation}

Using Eq.~\eqref{eq:H0_eigenform},  the time evolution of $H^\prime_I(\tau)$ can be obtained as
\begin{eqnarray}
    H^\prime_I(\tau)&=&e^{iH_0\tau}H^\prime e^{-iH_0\tau}\nonumber\\
    &=&\sum_{q,q^\prime,j,j^\prime} V^{j,j^\prime}_{q,q^\prime} \gamma^\dagger_{qj}\gamma_{q^\prime j^\prime}e^{i(2\epsilon_{qj}-2\epsilon_{q^\prime j^\prime})\tau}.
\end{eqnarray}
Moreover, for our perturbation 
$H^\prime={J_p}(ic_{m,2}c_{m+1,1}-ic_{m+1,1}c_{m,2})/8$
with $m$ at the center of the edge, $V^{j,j^\prime}_{k+q,k}$ can be computed as \begin{eqnarray}
V^{j,j^\prime}_{k+q,q}&=&i\frac{J_p}{4N_y}(e^{iq\delta_z-ikr_{m}}\hat{f}^{2\ast}_{k+q,j}\hat{f}^{1}_{q,j^\prime}\nonumber \\ &&-e^{-iq\delta_z -ik r_{m+1}}\hat{f}^{2}_{q j^\prime}\hat{f}^{1\ast}_{k+q j}).
\end{eqnarray}
Here, $\hat{f}^{l\ast}_{qj}$ is the eigenmode component of the Majorana fermion in momentum space, i.e., $c^\dagger_{ql}=\sum_j \hat{f}^{l\ast}_{qj}\gamma^\dagger_{qj}$. 

In addition, the operator $e^{iH_0t}c^\dagger_{q}c_{k+q}e^{-iH_0t}$ in Eq.~(\ref{eq:Ck_linear}) can be evaluated to give
\begin{eqnarray}
    &&e^{iH_0t}i (c^\dagger_{q,2}c_{k+q,1}e^{i(k+q)\delta_z}-c^\dagger_{q,1}c_{k+q,2}e^{-iq\delta_z})e^{-iH_0t}\nonumber \\   
    &=&i\sum_{jj^\prime} (\hat{f}^{2\ast}_{qj}\hat{f}^{1}_{k+q,j^\prime}e^{i(k+q)\delta_z}
-\hat{f}^{1\ast}_{qj}\hat{f}^{2}_{k+q,j^\prime}e^{-iq\delta_z})\nonumber \\ &&\times\gamma^\dagger_{qj}\gamma_{k+q,j^\prime}e^{2i(\epsilon_{qj}-\epsilon_{k+q,j^\prime})t},\\
&\equiv&\sum_{jj^\prime}F^{jj^\prime}_{q,k+q}\gamma^\dagger_{qj}\gamma_{k+q,j^\prime}e^{2i(\epsilon_{qj}-\epsilon_{k+q,j^\prime})t}.
\end{eqnarray}

Then, using the identity ${\rm Tr}( \gamma^\dagger_{kj}\gamma_{k^\prime j^\prime}\rho_0)=\delta_{jj'}\delta_{kk'}$ for $j,j'<0$, we obtain 
\begin{eqnarray}
    C(k,t)&=&\frac{J}{2}\sum_{q}\sum^N_{j^\prime=1}\sum_{j=-N}^{-1} F^{jj^\prime}_{q,k+q} \frac{e^{2i(\epsilon_{qj}-\epsilon_{k+q,j^\prime})t}}{-2\epsilon_{qj}+2\epsilon_{k+q,j^\prime}} V^{j^{\prime},j}_{k+q,q}\nonumber \\ &&
    -F^{j^\prime j}_{q,k+q} \frac{e^{2i(\epsilon_{qj^\prime}-\epsilon_{k+q,j})t}}{2\epsilon_{k+q,j}-2\epsilon_{qj^{\prime}}} V^{j,j^{\prime}}_{k+q,q},
     \label{eq:Cqt}
\end{eqnarray}   
and we dropped the time-independent terms along the way.

\begin{figure}[t!]
\includegraphics[width = 0.98\linewidth]{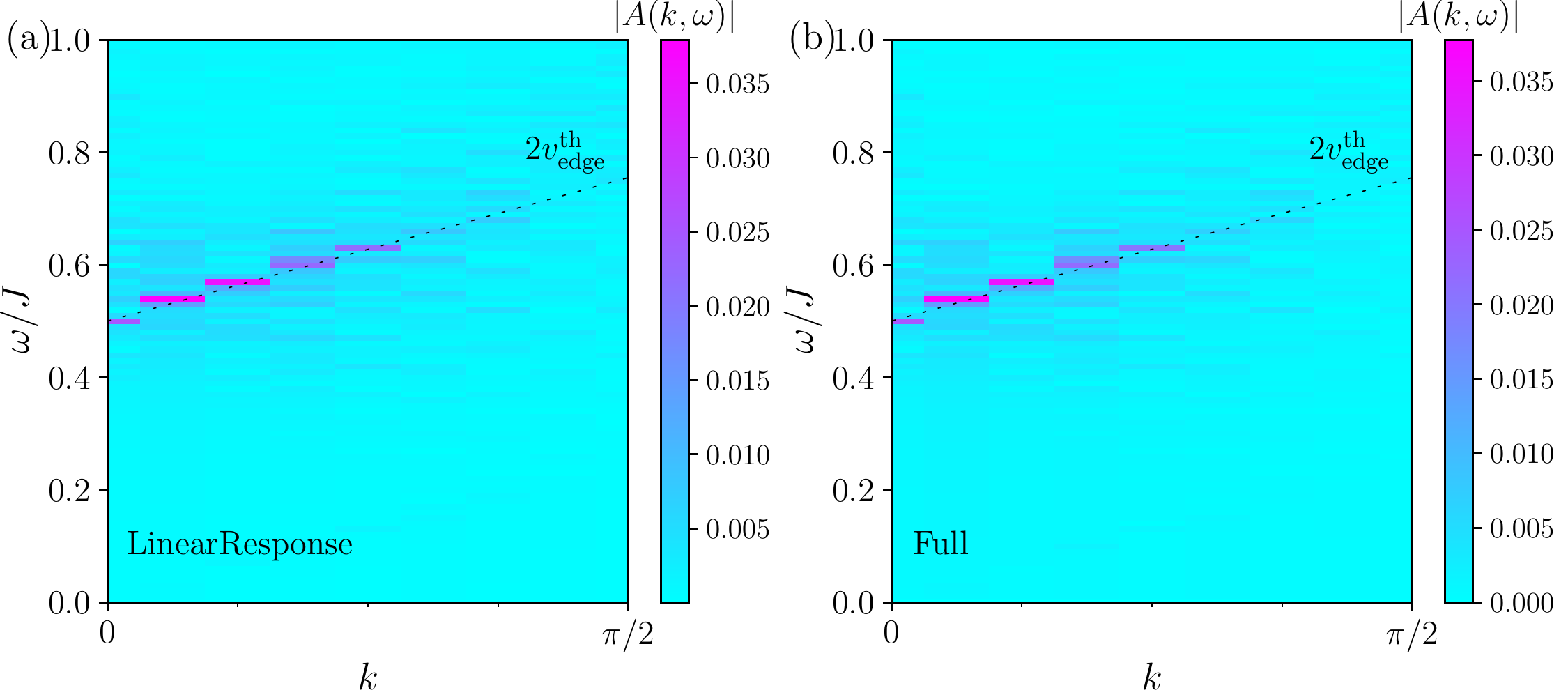}
\caption{Fourier spectrum of the $zz$-correlation calculated from (a) the linear response method, and (b) the exact numerical simulation. The system size is $16_x\times 32_y$, and the model parameters are $\omega_D=8/J$, $J_q/J=0.008$, and $N_T=800$. The black dashed line
corresponds to the theory prediction: $2v^{\rm th}_{\rm edge}k+1/2$, see Eq.~\eqref{resultsDelta}. }
\label{figLR}
\end{figure}

From Eq.~\eqref{eq:Cqt}, one can calculate the Fourier spectrum by using Eq.~\eqref{eq:FourierTrans} with $\ell=t/T$. We can then compare the resulting expression to our exact numerical simulation, shown in Fig.~\ref{fig_3}(b). The results are shown in Fig.~\ref{figLR}, showing an almost perfect agreement. 
This allows us to analytically analyze the dominant signal, from which we extract the edge velocity. 
To this end, we employ the delta-function identity $\int_0^\infty e^{i\omega t} dt=\pi \delta(\omega)$ to calculate the Fourier spectrum, in the limit $t\to\infty$. This gives
\begin{eqnarray}
    &&C(k,\omega)=\int_0^\infty \mathrm dt\; C(k,t) e^{i\omega t} \nonumber \\  
    &=&\frac{J\pi}{2}\sum_{q}\sum^N_{j^\prime=1}\sum_{j=-N}^{-1} F^{jj^\prime}_{q,k+q} V^{j^{\prime},j}_{k+q,q} \frac{\delta({2\epsilon_{q,j}-2\epsilon_{k+q,j^\prime}}+\omega)}{-2\epsilon_{q,j}+2\epsilon_{k+q,j^\prime}}\nonumber \\
    &&-F^{j^\prime j}_{q,k+q}\frac{\delta(2\epsilon_{q,j^\prime}-2\epsilon_{k+q,j}+\omega)}{2\epsilon_{k+q,j}-2\epsilon_{q,j^{\prime}}} V^{j,j^{\prime}}_{k+q,q}.
\end{eqnarray}

We recall from Fig.~\ref{fig_2}(a) in the main text, that when $q_y=\pi$, the eigenstates are highly degenerate and have energy $\pm J/4$ (crossing point of all curves). 
There are $N_x-1$ degenerate energy levels in each of the positive and negative dispersion branches. 
%The degenerate number for each energy is $N_x-1$ with the corresponding state indices being 2 to $N_x$ and $-2$ to $-N_x$, respectively. 
Hence, $C(k,\omega)$ will be dominated by those values of $\omega$ that match the energy difference in the argument of the delta functions, so that all degenerate states contribute. 
To account for this degenerate contribution, one can take 
one of the two states in the delta function to be at the degeneracy point (i.e., $\epsilon_{q=\pi,\pm j}=\pm J/4$, with $j>1$), and the other at the edge state (i.e., $\epsilon_{k+q=\pi+k,\pm 1}=\pm v_{\rm edge} k$). Then, the two delta functions in the equation above become 
\begin{eqnarray}
\delta({2\epsilon_{q,j}-2\epsilon_{k+q,j^\prime}}+\omega)&=& \delta\left(\frac{1}{2}J+2v_{\rm edge} k - \omega\right),\nonumber\\
\delta(2\epsilon_{q,j^\prime}-2\epsilon_{k+q,j}+\omega)
&=& \delta\left(\frac{1}{2}J+2v_{\rm edge} k + \omega\right).\notag \\
\label{resultsDelta}
\end{eqnarray}
We verified that this result matches our numerical simulation; see~Fig.~\ref{figLR}. In particular, we conclude that the correlation between the highly degenerate state and the edge state dominates the signal $|A(k,\omega)|$. 

Importantly, the result in Eq.~\eqref{resultsDelta} shows that the dominant spectroscopic signal can be used to probe the dispersion relation
$$\omega = (1/2)J + 2 v_{\rm edge} k,$$ and hence, to extract the velocity of the edge mode $v_{\rm edge}$, as we now describe in the following section.

\section{\label{app:fitting_procedure}Extracting the edge mode velocity}

Finally, we turn our attention to the fitting procedure that we use to extract (twice) the chiral edge mode velocity $v_\mathrm{edge}$ from the Fourier spectrum defined in Eq.~\eqref{eq:FourierTrans}.

Consider the Fourier spectrum as a dataset of triples
\begin{equation}
    \label{eq:data}
    \mathcal{D} = \{\left(k_y^j,\omega_\ell, |A(k_y^j,\omega_\ell)|\right)\},
\end{equation}
where to each Fourier momentum and frequency point, we associate the corresponding value of $|A(k_y^j,\omega_\ell)|$.
Our goal is to extract the slope of the most dominant part of the Fourier signal, which corresponds to the discernible chiral pattern. Thus, in constructing the dataset $\mathcal{D}$, in order to focus the signal on the dominant chiral region, we only consider points $(k_y^j,\omega_\ell)$ for which the strength of the Fourier spectrum is larger than $40\%$ of the maximum available signal $|A(k_y^j,\omega_\ell)|$.

Comparing this problem to ordinary linear regression, here we have to take into account the strength of the Fourier spectrum. Thus, we apply a weighted linear regression, where each $(k_y^j,\omega_\ell)$ point is additionally multiplied by its strength $|A(k_y,\omega)|$. The corresponding cost function reads as 
\begin{widetext}
\begin{equation}
    \label{eq:fit_cost}
    \mathcal{L}(v_\mathrm{edge}) = \sum_{(k^j_y,\omega_\ell)\in\mathcal{D}}
    \frac{
    |A(k^j_y,\omega_\ell)|
    }
    {\sum_{(k^j_y,\omega_\ell)\in\mathcal{D}}|A(k_y^j,\omega_\ell)|}
    \left(v_\mathrm{edge} k^j_y + \frac{1}{2}J - \omega_\ell\right)^2\; .
\end{equation}
\end{widetext}
This cost function $\mathcal{L}$ can be interpreted as the  variance or error of the fit. Thus, we can use the value $\mathcal{L}(v_\mathrm{edge})$ to define the error bars on the extracted value for $v_\mathrm{edge}$.
Note that, in principle, one can also leave the $\omega$-axis intercept as a fitting parameter; instead, we use the value $1/2$ which is obtained from the linear response analysis (cf.~Eq.~\eqref{resultsDelta}); we also verified this independently from the fit in the clean theoretical regime of large $N_T$ and $t_\mathrm{ramp}$. By adopting this value, we reduced the fitting process to only one parameter, namely, the slope. Thus, a limited number of critical data points are sufficient for the fitting, and these can be obtained by selecting appropriate hyperparameters; the procedure maintains a high degree of robustness as long as the cutoff is sufficiently high and the introduced noisy data points are negligible.

\begin{figure}[t!]
\includegraphics[width = 0.95\linewidth]{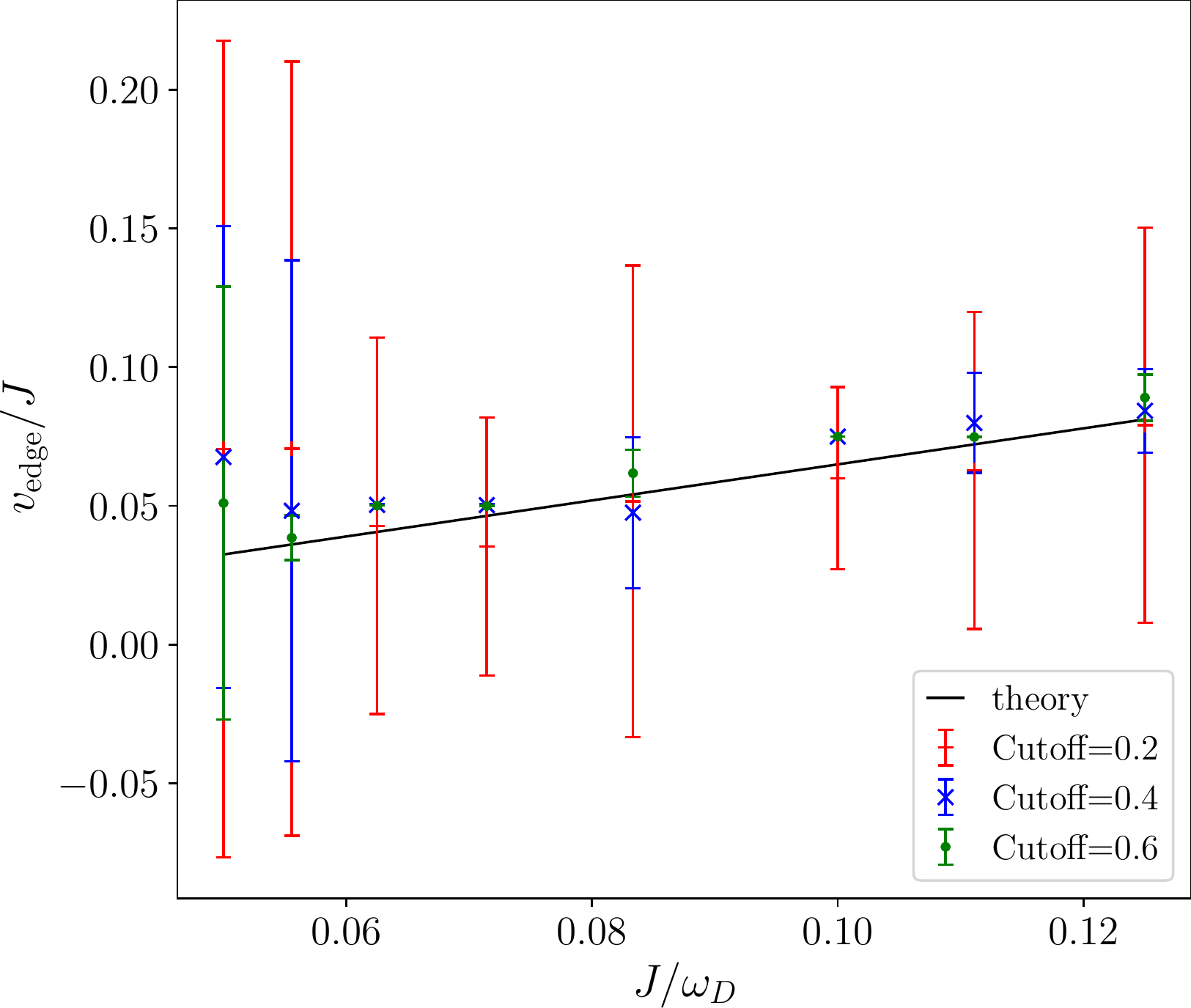}
\caption{
Edge state velocity $v_\mathrm{edge}$ extracted from fits to $|A(k,\omega)|$ with different cutoff values at $Jt_\mathrm{ramp}=250$. For a small cutoff value of $0.2$, residual excitations dominate the Fourier signal which leads to larger error bars. Additionally, for larger values of $\omega_D/J$, the gap is smaller, and the strength of the chiral signal is comparable to the strength of residual excitations due to the finite ramp and observation times, which also results in larger error bars even if the cutoff value is relatively large. Here, $N_T=50\omega_D/J+1$ is chosen long enough to resolve the chiral signal. The system size is $8_x\times16_y$ (cylinder geometry).
}
\label{figA7}
\end{figure}

Figure~\ref{figA7} shows the dependence of the extracted edge velocity on the drive frequency $\omega_D$ with different cutoffs; we use the scaling $N_T=50\omega_D/J+1$ to keep the total physical observation time $t_\mathrm{obs}=TN_T$ approximately the same throughout the drive frequency axis, and $Jt_\mathrm{ramp}{=}250$. By using these parameters and ignoring the $k=0$ point, the fitting procedure shows a decent resolution when the chiral gap and the cutoff are sufficiently large (i.e., for $J/\omega_D>0.06$ and cutoff value not less than $0.4$). However, when the gap or the cutoff value is small, the effect of bulk states excited during the ramp cannot be disregarded these excitation points then lead to large error bars as exhibited by the $J/\omega_D<0.06$ region or the cutoff value of $0.2$ (red data points in the figure). 
Thus, potential experiments should target the parameter regime with the largest accessible chiral gap and choose a relatively large cutoff in the fitting.

\bibliography{./floquet_kitaev.bib}

\end{document}